%% file: main.tex
\documentclass[sigconf,screen,dvipsnames]{acmart}

\AtBeginDocument{%
  \providecommand\BibTeX{{%
    \normalfont B\kern-0.5em{\scshape i\kern-0.25em b}\kern-0.8em\TeX}}}

\copyrightyear{2021}
\acmYear{2021}
\setcopyright{acmlicensed}\acmConference[MICRO '21]{MICRO'21: 54th Annual IEEE/ACM International Symposium on Microarchitecture}{October 18--22, 2021}{Virtual Event, Greece}
\acmBooktitle{MICRO'21: 54th Annual IEEE/ACM International Symposium on Microarchitecture (MICRO '21), October 18--22, 2021, Virtual Event, Greece}
\acmPrice{15.00}
\acmDOI{10.1145/3466752.3480110}
\acmISBN{978-1-4503-8557-2/21/10}

\usepackage{xspace}
\usepackage{subcaption}
\usepackage{pifont}
\usepackage[us,12hr]{datetime}

\usepackage{makecell}
\usepackage{booktabs}
\usepackage{multirow}
\usepackage{multicol}
\usepackage{colortbl}
\usepackage[compact]{titlesec} 
\usepackage{siunitx}
\usepackage{stfloats}
\usepackage{setspace}

\newcommand{\cmark}{\ding{51}}%
\newcommand{\xmark}{\ding{55}}%

\newcommand{\squishlist} {
    \begin{list}{$\bullet$} {
        \setlength{\itemsep}{-1pt}
        \setlength{\parsep}{2pt}
        \setlength{\topsep}{0pt}
        \setlength{\partopsep}{0pt}
        \setlength{\leftmargin}{1.0em}
        \setlength{\labelwidth}{1em}
        \setlength{\labelsep}{0.5em}
    }
}

\newcommand{\squishend} {
    \end{list}
}

\newcommand{\mr}[2]{\multicolumn{1}{c}{\multirow{#1}{*}{\makecell{#2}}}}
\newcommand{\stripe}{\rowcolor{blue!5}}

\newcommand{\incircle}[1]{%
    \IfEqCase{#1}{%
        {1}{\ding{182}}%
        {2}{\ding{183}}%
        {3}{\ding{184}}%
        {4}{\ding{185}}%
        {5}{\ding{186}}%
        {6}{\ding{187}}%
        {7}{\ding{188}}%
        {8}{\ding{189}}%
    }[\PackageError{mycirc}{Undefined option to mycirc: #1}{}]%
}

\input{sections/vars}

\newcommand{\methodunformatted}{U-TRR\xspace}
\newcommand{\method}{\texttt{\methodunformatted{}}\xspace}

\newcommand{\rtplong}{\emph{{\hhf{Row Scout}}}\xspace}
\newcommand{\rtp}{\emph{{\hhf{RS}}}\xspace}
\newcommand{\trranlong}{\emph{{TRR Analyzer}}\xspace}
\newcommand{\trran}{\emph{{\mbox{TRR-A}}}\xspace}

\newcommand{\ignore}[1]{}

\newcommand{\tras}{\texttt{{tRAS}}\xspace}
\newcommand{\trp}{\texttt{{tRP}}\xspace}

\newcommand{\tfaw}{\texttt{{tFAW}}\xspace}

\newcommand{\cmdact}{\texttt{{ACT}}\xspace}
\newcommand{\cmdread}{\texttt{{RD}}\xspace}
\newcommand{\cmdwrite}{\texttt{{WR}}\xspace}
\newcommand{\cmdprech}{\texttt{{PRE}}\xspace}
\newcommand{\cmdrefresh}{\texttt{{REF}}\xspace}

\newtheorem{obsA}{\hspace*{-1em}\textbf{Vendor A | Observation}}
\newtheorem{obsB}{\hspace*{-1em}\textbf{Vendor B | Observation}}
\newtheorem{obsC}{\hspace*{-1em}\textbf{Vendor C | Observation}}

\newtheorem{req}{Requirement}

\newif\ifcameraready
\camerareadytrue

\newcommand{\versionnum}[0]{5.7}

\fancypagestyle{firstpage}{
  \fancyhf{}
  
  \ifcameraready
    \fancyfoot[C]{\thepage}
    \pagenumbering{arabic}
  \else
    \fancyhead[C]{
    {\textcolor{blue}{\emph{Version \versionnum~---~\today, \ampmtime}}}} 
    \fancyfoot[C]{\thepage}
    \pagenumbering{arabic}
  \fi
}

\ifcameraready
    
    \newcommand{\hh}[1]{{#1}\xspace}
    \newcommand{\hht}[1]{{#1}\xspace}
    \newcommand{\hhth}[1]{{#1}\xspace}
    \newcommand{\hhf}[1]{{#1}\xspace}
    \newcommand{\hhs}[1]{{#1}\xspace}
    \newcommand{\hhe}[1]{{#1}\xspace}
    \newcommand{\hhn}[1]{{#1}\xspace}
    
    \newcommand{\jk}[1]{{{#1}}\xspace}
\else

    \newcommand{\hh}[1]{{#1}\xspace}
    \newcommand{\hht}[1]{{#1}\xspace}
    \newcommand{\hhth}[1]{{#1}\xspace}
    \newcommand{\hhf}[1]{{#1}\xspace}
    \newcommand{\hhs}[1]{{#1}\xspace}
    \newcommand{\hhe}[1]{{#1}\xspace}
    \newcommand{\hhn}[1]{{\color{MidnightBlue}{#1}}\xspace}
    
    \newcommand{\jk}[1]{{#1}\xspace}
\fi

\newcommand{\authspace}[0]{\hspace{32pt}}
\newcommand{\affilspace}[0]{\hspace{22pt}}

\newcommand{\affilETH}[0]{\textsuperscript{$\dag$}}
\newcommand{\affilTOBB}[0]{\textsuperscript{$\ddag$}}
\newcommand{\affilQualcomm}[0]{\textsuperscript{$\sigma$}}

\usepackage{etoolbox}

\begin{document}

\title{Uncovering \hhs{In-DRAM} RowHammer Protection Mechanisms:\\A New Methodology, Custom RowHammer Patterns, and Implications}

\makeatletter
\patchcmd{\@titlefont}{\Huge\sffamily\bfseries}{\huge\sffamily\bfseries}{}{}
\patchcmd{\@subsubsecfont}{\sffamily\itshape}{\sffamily\bfseries}{}{}
\makeatother

\author{\vspace{-3pt}Hasan Hassan\affilETH \authspace Yahya Can Tu\u{g}rul\affilETH\affilTOBB \authspace Jeremie S. Kim\affilETH \authspace
                        Victor van der Veen\affilQualcomm \authspace \\ Kaveh Razavi\affilETH \authspace Onur Mutlu\affilETH}
\affiliation{\vspace{0pt}\affilETH \textit{ETH Z{\"u}rich} \affilspace \affilTOBB
\textit{TOBB University of Economics \& Technology} \affilspace \affilQualcomm \textit{Qualcomm Technologies Inc.} \ \\\vspace{2pt}\country{}}

\renewcommand{\shortauthors}{Hassan et al.}
\renewcommand{\shorttitle}{\methodunformatted{}: Uncovering and Exploiting in-DRAM RowHammer Protection Mechanisms}

\renewcommand{\authors}{Hasan Hassan, Yahya Can Tugrul, Jeremie S. Kim, Victor van der Veen, Kaveh Razavi, and Onur Mutlu}

\input{sections/0_abstract.tex}



\maketitle

\thispagestyle{firstpage}
\pagestyle{firstpage}

\input{sections/1_intro}
\input{sections/2_background}
\input{sections/3_overview}

\input{sections/4_retention_profiling}

\input{sections/5_analyzing_refs}
\input{sections/6_insights}
\input{sections/7_case_studies}
\input{sections/8_related_work}
\input{sections/9_conclusion}

\vspace{-1mm}
\begin{acks}
\vspace{-2mm}
\hht{We thank the anonymous reviewers of MICRO 2021 for feedback. We thank the
SAFARI Research Group members for valuable feedback and the stimulating
intellectual environment they provide. We acknowledge the generous gifts
provided by our industrial partners\hhs{, especially} Google, Huawei, Intel,
Microsoft, and VMware.} \hhth{This work was \hhf{also} supported \hhf{in part}
by the Netherlands Organisation for Scientific Research through grant NWO
016.Veni.192.262.}
\end{acks}

\bibliographystyle{IEEEtranS}
\bibliography{refs}

\end{document}

%% file: sections/vars.tex
\newcommand{\numTestedDIMMs}{45} 
\newcommand{\numTestedDIMMsFromB}{15} 
\newcommand{\vulnRowsMaxPct}{99.9} 
\newcommand{\maxPerBankBitflips}{9.4 million} 

%% file: sections/0_abstract.tex
\begin{abstract}

    The RowHammer vulnerability in DRAM is a critical threat to system security.
    To protect against RowHammer, vendors commit to security-through-obscurity:
    \hht{modern} DRAM \hhth{chips rely} on undocumented, proprietary, on-die
    mitigations, commonly known as \emph{Target Row Refresh} \hhn{(TRR)}.
    \hhth{At a high level, TRR detects and refreshes potential RowHammer-victim
    rows, but its exact implementations are not openly disclosed. Security
    guarantees of \hhn{TRR mechanisms} cannot be easily studied due to their
    proprietary nature.}

    To assess \hh{the} security guarantees of recent \hh{DRAM} \hhth{chips}, we
    present \emph{\hh{Uncovering} TRR} (\method{}), an experimental methodology
    to analyze \hht{in-DRAM} TRR implementations. \method{} is based on the \hhth{new}
    observation that \hht{data} retention failures in DRAM enable a side channel
    that leaks information on how TRR refreshes potential victim rows.
    \method{} allows us to (i) understand how logical \hh{DRAM} rows are laid
    out physically in silicon; (ii) study undocumented on-die TRR
    \hh{mechanisms}; and (iii) combine (i) and (ii) to evaluate \hh{the}
    RowHammer security guarantees of modern DRAM \hht{chips}. We show how
    \method{} allows us to craft RowHammer access patterns that successfully
    circumvent \hhth{the TRR mechanisms \hhf{employed in \numTestedDIMMs{} DRAM
    modules} of the three major} DRAM vendors. We find that \hhe{the
    \hhf{DRAM} modules we} \hhf{analyze} are vulnerable to RowHammer\hht{,
    having} bit flips in up to \vulnRowsMaxPct{}\% of all DRAM rows.
    We make \method{} source code openly and freely available at~\cite{utrrsource}.
    
\end{abstract}

%% file: sections/1_intro.tex

\vspace{-2mm}

\section{Introduction}
\label{sec:intro}
\vspace{-1mm}

DRAM~\cite{dennard1968field} has long been the dominant memory technology
used in almost all computing systems due to its low latency and low cost
per bit. DRAM vendors still push DRAM technology scaling forward to
continuously shrink DRAM cells to further reduce the cost of
DRAM~\cite{mutlu2013memory}. Unfortunately, DRAM technology scaling
leads to increasingly important DRAM reliability problems as DRAM cells get
smaller and the distance between the cell reduce. Kim et
al.~\cite{kim2014flipping} report that most DRAM chips manufactured
since 2010 are susceptible to disturbance errors that are
popularly referred to as RowHammer. Recent works~\cite{yang2019trap,
walker2021dram, kim2014flipping, mutlu2019rowhammer, ryu2017overcoming,
park2016experiments, yang2016suppression, yang2017scanning, gautam2019row,
jiang2021quantifying} explain the circuit-level
charge leakage mechanisms that lead to the RowHammer vulnerability. 
A security attack exploits the RowHammer vulnerability by hammering
(i.e., repeatedly activating and precharging) an \emph{aggressor} row
many times (e.g., $139K$ in DDR3~\cite{kim2014flipping}, $10K$ in
DDR4~\cite{kim2020revisitingrh}, and $4.8K$ in
LPDDR4~\cite{kim2020revisitingrh})\footnote{For DDR3 chips,
\cite{kim2014flipping} reports the minimum number of row activations on a
\emph{single} aggressor row (i.e., single-sided RowHammer) to cause a RowHammer
bit flip. For DDR4 and LPDDR4 chips, \cite{kim2020revisitingrh} reports the
minimum number of row activations to \emph{each of the two}
immediately-adjacent aggressor rows (i.e., double-sided RowHammer).}
to cause bit flips in the cells of the \emph{victim} rows that are
physically adjacent to the hammered row. Since the discovery of RowHammer,
researchers have proposed many techniques that take advantage of the
RowHammer vulnerability to compromise operating
systems~\cite{seaborn2015exploiting, van2018guardion, van2016drammer,
gruss2016rowhammer, cojocar2019exploiting, qiao2016new, kwong2020rambleed,
pessl2016drama, bhattacharya2016curious, jang2017sgxbomb, zhang2020pthammer,
cojocar2020we, weissman2020jackhammer, ji2019pinpoint}, web
browsers~\cite{gruss2018another, bosman2016dedup, frigo2020trrespass,
deridder2021smash, frigo2018grand}, cloud virtual machines~\cite{razavi2016flip,
xiao2016one}, remote servers~\cite{tatar2018defeating, lipp2018nethammer}, and
deep neural networks~\cite{yao2020deephammer,
hong2019terminal}.\footnote{A review of many RowHammer attacks and
defenses is provided in~\cite{mutlu2019rowhammer}.} Thus, RowHammer is
a clear major threat to system security.

To prevent RowHammer, DRAM vendors equip their DRAM chips with a
\emph{mitigation mechanism} known as Target Row Refresh
(TRR)~\cite{frigo2020trrespass, cojocar2019exploiting,kim2020revisitingrh,
micronddr4trr}. The main idea of TRR is to detect an aggressor row
(i.e., a row that is being rapidly activated) and refresh its
victim rows (i.e., neighboring rows that are physically
adjacent to the aggressor row). TRR refreshes the victim rows
separately from the \emph{regular} refresh operations~\cite{raidr,
venkatesan2006retention, liu2013experimental, cui2014dtail, bhati2015flexible}
that must periodically (e.g., once every \SI{64}{\milli\second}) refresh
each DRAM row in the entire DRAM chip. Some of the major DRAM vendors
advertise \emph{RowHammer-free} DDR4 DRAM
chips~\cite{samsung2014trrDram, micronddr4, frigo2020trrespass}.
However, none of the DRAM vendors have so far disclosed the implementation
details let alone proved the protection guarantees of their TRR
mechanisms. It is long understood that security cannot be achieved
only through obscurity~\cite{scarfone2008guide, anderson2020security,
saltzer1975protection}. Yet, such is the current state of practice when it comes
to DRAM.

The recent TRRespass work~\cite{frigo2020trrespass} shows that in certain DRAM
chips, the TRR mechanism keeps track of only a few aggressor
rows. Hence, an access pattern that hammers many aggressor rows can
circumvent the TRR protection and induce RowHammer bit
flips.
While TRRespass already demonstrates
the flaws of certain TRR implementations, it does \emph{not} propose
a methodology for \emph{discovering} these flaws. According
to~\cite{frigo2020trrespass}, simply increasing the number of aggressor
rows is \emph{not} sufficient to induce bit flips on 29 out
of 42 of the DDR4 modules that were tested by~\cite{frigo2020trrespass}.
However, it is unclear whether such DDR4 chips are \emph{fully}
secure due to the lack of insight into the detailed operation of TRR in these
chips. Thus, we need new methods for identifying whether or not
DRAM chips are fully secure against RowHammer.

\vspace{1mm}
\noindent \textbf{\methodunformatted.} We develop \method{}, a \hhth{new and}
practical methodology that \hhf{uncovers} the inner workings of TRR mechanisms
in modern DRAM \hht{chips}. \hhth{To develop \method{}, we formulate} TRR as a
function that takes a number of DRAM accesses including \hhth{aggressor rows} as
input and refreshes a number of rows that are detected as victims. \hht{The goal
of} \method{} \hhth{is to enable the observation (i.e., uncovering) of} all
refreshes generated by TRR after \hhth{inducing a carefully crafted sequence of
DRAM accesses}. To make this possible, we make a key observation that retention
failures \hhf{that occur} on a DRAM row can be used as a side channel to detect
\hh{\emph{when} the row is} refreshed, due to \hht{either} TRR or \hh{periodic
refresh operations}. \hhth{We easily distinguish TRR-induced refreshes from
regular refreshes as the latter occur at fixed time intervals independently of
the access pattern.} 

We build two new tools, \rtplong{} \hhf{(\rtp{})} and \trranlong{}
\hhs{(\trran{})}, that make use of this \hhth{new} observation \hhth{to
construct \method{} and thus enable} a deep analysis of TRR. \hhf{The goal of
\rtp{} is to find a set of DRAM rows \hhf{that meet certain requirements as
needed by a \trran{} experiment} and identify the data retention times of these
rows. The goal of \trran{} is to use the \rtp{}-provided rows to determine when
a TRR mechanism refreshes a victim row by exactly distinguishing between TRR
refreshes and regular refreshes, and thus build an understanding of the
underlying TRR operation.}

\hht{\rtplong{} (\rtp{})} \hhth{profiles} \hht{the \hhth{data} retention time of
DRAM rows and passes to \trran{} a list of rows that satisfy two key
requirements}. First, a DRAM row should have a \emph{consistent} retention time
\hht{that does \emph{not}} vary over time based on effects such as Variable
Retention Time (VRT)~\hht{\cite{qureshi2015avatar, yaney1987meta,
liu2013experimental, restle1992dram}}. \hht{\rtp{} should not \hhf{supply to
\trran{}} a row with an inconsistent retention time since it would not be
possible for \trran{} to distinguish whether the row has been refreshed or it
simply retained its data correctly for longer than the profiled retention time.}
Second, \hhth{to enable observing how many and which DRAM rows TRR treats as
victim rows,} \hht{\rtp{} should provide \emph{multiple}} DRAM rows \hht{that}
have similar retention times and \hhf{that are} located at \hht{certain
\emph{configurable}} distances with respect to each other. \hhth{It is crucial
to find rows with similar retention times in order to observe whether or not TRR
can refresh multiple rows at the same time.} These two requirements enable
reliable and precise analysis \hhth{of TRR-induced refreshes} by \trran{}. 

\hhth{\hhs{\trranlong{} (\trran{})} discovers 1) access patterns that
\hhf{cause} the TRR mechanism to treat a certain row as an aggressor row and 2)
when a TRR-induced refresh targets a victim row.}
\hhf{At a high level, \jk{\trran{} infers the occurrence of a TRR-induced
refresh operation to an \rtp{}-provided row {if the row does \emph{not} contain
any bit flips \hhf{after disabling regular refreshes and accesses to the row for
its \rtp{}-profiled retention time.} Only a TRR-induced refresh can prevent the
\rtp{}-provided row from experiencing bit flips} by refreshing the row before it
experiences a retention failure.} Thus, \trran{} attributes \emph{not} observing
a bit flip in an \rtp{}-provided row to TRR-induced refresh.}

\trran{} operates in three main steps. First, it initializes \hh{to known data
(e.g., all \emph{ones}) i) the \hht{\rtp{}-provided rows} and ii) the \hht{rows
\hhth{it selects} as aggressor rows in the experiment}}. Then, \trran{} lets the
\hht{\rtp{}-provided rows} leak their charge for half of their profiled
retention time. Second, \trran{} hammers (i.e., \hhf{repeatedly} activates and
precharges) the \hht{aggressor rows}. \hhf{After the hammers,} \trran{}
\hhf{issues} a small number of \hh{DRAM refresh commands (originally used for
only periodic refresh)} so \hhf{that} the \hhf{TRR-induced} row refresh can take
place.\footnote{\hhth{The current TRR implementations avoid changing the DDR
interface by piggybacking TRR-induced refreshes to regular refresh
commands~\cite{frigo2020trrespass, yauglikcci2021security}.}} \trran{} \jk{then}
waits for the \hhth{remaining} half of the profiled retention time.  \jk{If an
\rtp{}-provided row was not refreshed by TRR-induced refresh operations as a
result of the hammers issued by \trran{}, the \rtp{}-provided row would now have
not been refreshed for its full profiled retention time and will contain bit
flips.} 
Third, \trran{} reads back the data stored in the \hht{\rtp{}-provided rows}
and checks for bit flips. \hhf{\emph{Observing no bit flips}} \hhth{in an
\rtp{}-provided row} indicates that \hhth{either a TRR-induced or regular
refresh, which \trran{} can easily distinguish \hhf{between} since the regular
refreshes happen periodically,} targeted the \hhth{same row} during step two.
In contrast, \emph{observing \hhth{bit flips}} indicates that the
\hhth{\rtp{}-provided row was} never refreshed. These three steps constitute
the core of the experiments that we conduct to understand different in-DRAM TRR
implementations.

\vspace{1mm}
\noindent \textbf{Security analysis of TRR.} We use the \rtp{} and \trran{} on
\numTestedDIMMs{} DDR4 modules from the three major DRAM chip vendors (i.e.,
Micron, Samsung, SK Hynix). Our methodology \hhf{uncovers} important details of the
in-DRAM TRR designs of all vendors. We demonstrate the usefulness of these
insights \hhth{for} developing effective RowHammer attacks \hhth{on} DRAM
\hht{chips} from \hh{each vendor} by crafting specialized DRAM access patterns
that hammer \hht{a row} enough times to cause a RowHammer bit flip
\emph{without} alerting the TRR protection \hhth{mechanism} (i.e., \hhf{by}
redirecting \hhf{potential TRR} refreshes \emph{away from} the victim rows).
\hhth{In our evaluation, we find that all tested DRAM modules with different
manufacturing dates \hhf{(from 2016 to 2020)} are vulnerable to the new access
patterns we can craft via \method{}. Further, we find that \hhf{1) over
\vulnRowsMaxPct{}\% of the DRAM rows are vulnerable (i.e., have at least one bit
flip) to the new access patterns and 2) the new access patterns cause up to
\maxPerBankBitflips{} bit flips per DRAM bank.}} The large number of RowHammer
bit flips caused by our specialized \hhth{access} patterns has significant
implications for systems protected by \hhth{Error Correction Codes
(ECC)}~\cite{kang2014co,patel2020beer, patel2019understanding, patel2021harp}. Our analysis
shows that \hhf{the \method{}-discovered} access patterns can cause up to 7 bit
flips \hhf{at arbitrary locations} in one 8-byte dataword, suggesting that
typical ECC schemes capable of correcting one \hhth{error/symbol and detecting
two errors/symbols (e.g., SECDED ECC~\cite{micron2017whitepaper, oh2014a,
kwak2017a, kwon2017an, im2016im, son2015cidra, cha2017defect, jeong2020pair} and
Chipkill~\cite{dell1997white, nair2016xed, amd2009bkdg})} \hhf{\emph{cannot}}
provide sufficient protection against RowHammer \hhf{even in the presence of TRR
mechanisms}.

\vspace{1mm}
\noindent \textbf{Contributions.} We expect that \method will help future
research on \hhf{both} evaluating the security of existing RowHammer protections and the
design of more secure \hhth{RowHammer mitigation} \hh{mechanisms}. In summary,
we make the following \hhth{major} contributions:

\squishlist

    \item We develop \method{}, a \hhf{new} methodology for reverse-engineering
    \hhth{Target Row Refresh (TRR)} and regular refresh \hhf{mechanisms}.

    \item We use \method{} to \hhth{understand and uncover} the TRR
    implementations of \numTestedDIMMs{} DDR4 modules from the three \hhf{major
    DRAM} vendors. This evaluation shows that our \hhf{new} methodology is broadly
    applicable to any DRAM chip.

    \item Leveraging the TRR implementation details uncovered by \method{}, we
    craft specialized RowHammer access patterns that make \hhf{existing} TRR protections
    ineffective. 
    
    \item \hhth{Our specialized \hhf{\method{}-discovered} access patterns are
    significantly more effective than patterns from \hhth{the
    state-of-the-art}~\cite{frigo2020trrespass}: we show that our \hhs{new} RowHammer
    \hhth{access patterns cause} \hhf{1) bit flips \hhe{in \numTestedDIMMs{}} DDR4
    modules we comprehensively examine,} 2) bit flips in up to
    \vulnRowsMaxPct{}\% of the all rows in a DRAM bank, and 3) two and more (up
    to $7$) bit flips in a single 8-byte dataword, enabling practical RowHammer
    attacks in \hhth{systems that employ} ECC.}
        
\squishend

%% file: sections/2_background.tex
\section{Background}
\label{sec:background}
\vspace{-1mm}

We provide background on DRAM and the RowHammer phenomenon \hhth{that is}
required \hhth{for the reader} to understand how \method{} can precisely uncover
the behavior of an in-DRAM RowHammer mitigation mechanism. For \hhth{more
detailed descriptions} of DRAM organization and operation, we refer the reader
to \hhth{many} prior works~\cite{chang2016understanding, chang2014improving,
chang2016low, chang2017understanding, hassan2019crow, hassan2016chargecache,
kim2014flipping, kim2012case, lee2015adaptive, lee2013tiered,
lee2016simultaneous, lee-sigmetrics2017, lee2015decoupled, raidr,
liu2013experimental, seshadri2013rowclone, seshadri2017ambit,
seshadri2015gather, seshadri2020indram, zhang2014half, kim2020revisitingrh,
frigo2020trrespass, luo2020clrdram, yaglikci2021blockhammer, wang2020figaro,
cojocar2020we, patel2019understanding, bhati2015flexible, zhang2016restore,
kim2019d, kim2018dram, olgun2021quac, kim2016ramulator}.

\vspace{-1mm}
\subsection{DRAM Organization} 
\vspace{-1mm}

\begin{figure}[!h]
    \centering
    \vspace{-2mm}
    \includegraphics[width=.9\linewidth]{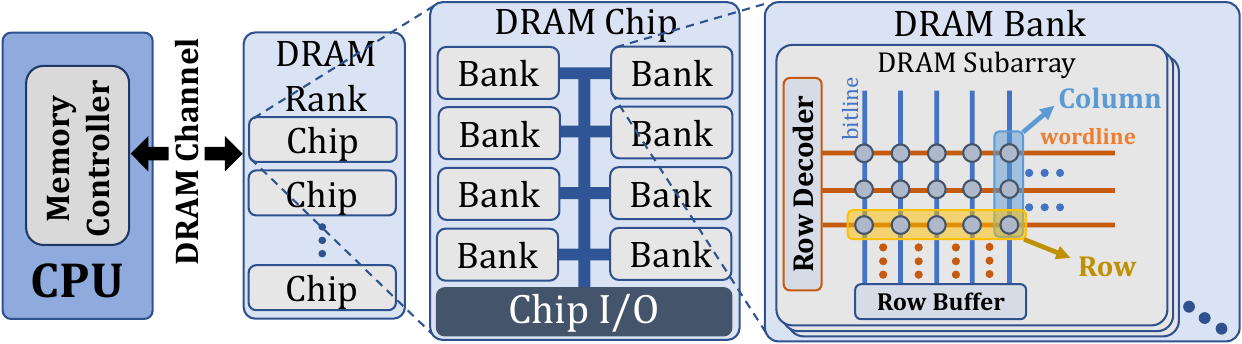}
    \vspace{-3mm}
    \caption{\hhn{Typical} DRAM \hhth{system} organization.}
    \vspace{-3mm}
    \label{fig:dram_organization}
\end{figure}

Fig.~\ref{fig:dram_organization} shows \hhf{the} typical organization of
\hhf{a} modern DRAM \hhth{system}. DRAM is organized into a hierarchical array
of billions of DRAM cells, each \hhth{holding} one bit of data. In modern
systems, a CPU chip implements a set of memory controllers, where each memory
controller interfaces with a DRAM \emph{channel} to perform read, write, and
maintenance operations (e.g., refresh) via a dedicated I/O bus that is
independent of other channels in the system. A DRAM channel can host one or
\hhf{more} \emph{DRAM modules}, where each module consists of \hhf{one or more
\emph{DRAM ranks.} A rank is comprised of multiple \emph{DRAM chips} that
operate in lock step and ranks \hhs{in} the same channel time-share the
channel's I/O bus.}

\vspace{-2mm}

A DRAM chip \hhth{consists of} multiple DRAM \emph{banks}, which share an
internal data/command bus.  Within a DRAM bank, DRAM cells are organized into
multiple (e.g., 128) dense two-dimensional arrays of DRAM cells called
\emph{subarrays}~\cite{kim2012case,chang2014improving,seshadri2013rowclone} and
corresponding peripheral circuitry for manipulating the data within the
subarray. \hhth{A} row of cells (i.e., DRAM \emph{row}) within a subarray share
a wire (i.e., \emph{wordline}), which \hhth{is} driven by a \emph{row decoder}
to \emph{open} (i.e., select) the row of cells to be read or written. \hhth{A}
column of cells (i.e., DRAM \emph{column}) within a subarray share a wire (i.e.,
\emph{bitline}), which is used to read and write to the cells with the help of a
\emph{row buffer} \hhth{(consisting of \emph{sense amplifiers})}. This
hierarchical layout of DRAM cells enables any data \hhth{in the DRAM system} to
be accessed \hhf{and updated} using unique \hhf{rank, bank,} row, and column
addresses. 

\subsection{DRAM Operation} 
\vspace{-1mm}

The memory controller interfaces with DRAM using a series of commands sent over
the I/O bus. To read or write data, the memory controller must first issue an
\emph{activate} (\cmdact{}) command to \emph{open} a row corresponding to the
\hhth{provided} memory address. When a row is activated, the data within the
DRAM row is copied into the row buffer \hhth{of the subarray}. The memory
controller then issues \emph{READ} (\cmdread{}) or \emph{WRITE} (\cmdwrite{})
commands to read or update the data \hhth{in} the row buffer. Changes to the
data in the row buffer propagate to the DRAM cells in the opened row. When data
\hhth{from} another row is required, the memory controller must first issue a
\emph{precharge} (\cmdprech{}) command \hhth{that} \emph{closes} the \hhth{open
row and prepares} the bank for \hhth{the activation of} another row. 

\noindent\textbf{DRAM Refresh.} \hhth{A DRAM cell stores \jk{a} data value} in
the form of charge in \hhth{its capacitor (e.g., a charged cell can represent 0
or 1 and vice versa). Since the capacitor naturally loses} charge over time,
\hhth{the capacitor charge must be actively and periodically \emph{refreshed}}
to prevent information loss \hht{due to a data retention
failure}~\cite{patel2017reaper, 2014lpddr4, jedec2012, liu2013experimental,
khan2014efficacy, khan2017detecting, khan2016parbor, raidr}. \hhth{To enable
such periodic refresh of all DRAM cells, } the memory controller must
\hhth{periodically} issue a \emph{refresh} (\cmdrefresh{}) command (\hhth{e.g.,}
every \SI{7.8}{\micro\second}) to ensure that every DRAM cell is refreshed once
\hhth{at a fixed \emph{refresh interval} (i.e., typically 32 or
\SI{64}{\milli\second})}~\cite{2014lpddr4, jedec2012, liu2013experimental,
raidr}.

\vspace{-1mm}
\subsection{RowHammer} 
\vspace{-1mm}

Modern DRAM \hht{chips} suffer from disturbance errors that occur when a high
\hhth{number} of activations \hhth{(within a refresh interval)} to one DRAM row
unintentionally affects the values of cells in nearby
rows~\cite{kim2014flipping}. This phenomenon, \hhf{popularly} called
\emph{RowHammer}~\cite{kim2014flipping, mutlu2019rowhammer}, stems from
electromagnetic interference between circuit elements. \hhth{RowHammer becomes
exacerbated} as \hhf{manufacturing} process technology node size \hhth{(and
hence DRAM \hhf{cell size})} \hhf{shrinks} and circuit elements are placed
closer together~\cite{kim2020revisitingrh,mutlu2017rowhammer}. As demonstrated
in prior work~\cite{kim2014flipping, kim2020revisitingrh}, the RowHammer effect
is strongest between \hhth{immediately physically-adjacent} rows.
\hhn{RowHammer} bit flips are most likely to appear in \hhth{neighboring rows}
\hhn{physically} adjacent to a \hhf{\emph{hammered row}} \hhth{that is activated
many times} \hhf{(e.g., $139K$ in DDR3~\cite{kim2014flipping}, $10K$ in
DDR4~\cite{kim2020revisitingrh}, and $4.8K$ in
LPDDR4~\cite{kim2020revisitingrh})}\footnotemark[1]. \hhf{A hammered row is also
called an \emph{aggressor row} and a nearby row \hht{that is} affected by
\hht{the hammered row is called}a \emph{victim row}, regardless of whether or
not the victim row \hhf{actually experiences} RowHammer bit flips}.

\vspace{-3mm}
\begin{figure}[!h]
    \captionsetup[subfigure]{justification=centering, size=scriptsize}
    \centering
    \begin{subfigure}[b]{.48\linewidth}
        \centering
        \includegraphics[width=\linewidth]{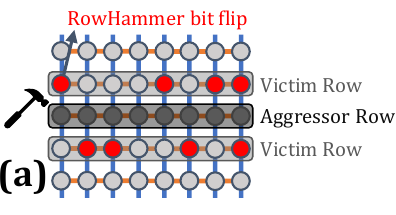}
    \end{subfigure}\quad
    \begin{subfigure}[b]{.48\linewidth}
        \centering
        \includegraphics[width=\linewidth]{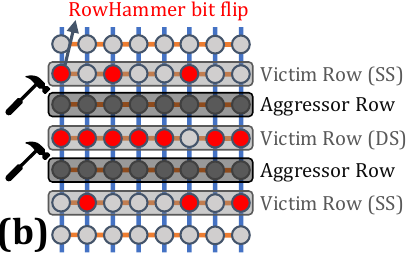}
    \end{subfigure}
    \vspace{-9mm}
    \caption{Typical Single-sided \hhf{(SS)} and Double-sided \hhf{(DS)} RowHammer access patterns.}
    \vspace{-4mm}
    \label{fig:rowhammer_access}
\end{figure}

To most effectively exploit the RowHammer phenomenon, attackers \hht{typically}
perform $i$) single-sided RowHammer (i.e., repeatedly \hhth{activate} one
\hhf{aggressor} row that is physically adjacent to the victim row, \hhth{as we
show in Fig.~\ref{fig:rowhammer_access}a})~\cite{kim2014flipping} or
\hht{$ii$)} double-sided RowHammer (i.e., repeatedly \hhth{activate in an
alternating manner} \hhf{two aggressor} rows that are \hhf{both} physically
adjacent to the victim row, \hhth{as we show in
Fig.~\ref{fig:rowhammer_access}b})~\cite{seaborn2015exploiting,
rh_project_zero}. \hhf{Prior works have shown that double-sided RowHammer leads
to more bit flips and does so more quickly than single-sided
RowHammer~\cite{kim2014flipping,kim2020revisitingrh, mutlu2019rowhammer,
seaborn2015exploiting, rh_project_zero}.}

\subsection{RowHammer Mitigation Mechanisms}
\label{subsec:rh_mitigation}
\vspace{-1mm}

\hh{To combat attacks that exploit the RowHammer phenomenon, various RowHammer
mitigation mechanisms have been proposed in
literature~\cite{yaglikci2021blockhammer, kim2014flipping, park2020graphene,
rh-apple, brasser2016can, konoth2018zebram, van2018guardion, aweke2016anvil,
lee2019twice, seyedzadeh2017counter, son2017making, you2019mrloc,
greenfield2016throttling, yauglikcci2021security, kim2021mithril,
taouil2021lightroad, devaux2021method}.\footnotemark[2]}
Yaglikci et al.~\cite{yaglikci2021blockhammer} classify these mitigation
mechanisms into four groups: $i$) increasing the refresh rate to reduce the
number of activations that can be performed within a refresh
interval~\cite{rh-apple, kim2014flipping}, $ii$) isolating sensitive data from
DRAM rows that an attacker can potentially hammer~\cite{brasser2016can,
konoth2018zebram, van2018guardion}, $iii$) keeping track of row activations and
refreshing potential victim rows~\cite{aweke2016anvil, kim2014flipping,
lee2019twice, park2020graphene, seyedzadeh2017counter, son2017making,
you2019mrloc, yauglikcci2021security, kim2021mithril, taouil2021lightroad,
devaux2021method}, and $iv$) throttling row activations to limit the times a row
can be activated within a refresh interval~\cite{yaglikci2021blockhammer,
kim2014flipping, greenfield2016throttling}.
Many of these \hhf{research} proposals describe the details of their proposed
mechanisms and discuss their security guarantees~\cite{kim2014flipping,
yaglikci2021blockhammer, park2020graphene}.

\hhf{Unfortunately}, DRAM vendors currently implement different
\emph{proprietary} in-DRAM RowHammer mitigation mechanisms, which they broadly
refer to as Target Row Refresh (TRR)~\cite{frigo2020trrespass,
cojocar2019exploiting,kim2020revisitingrh, micronddr4trr}. \hhn{TRR detects} a
potential aggressor row and \hhn{refreshes} its neighbor rows. The vendors have
\hhf{so far not} disclosed the implementation details of their TRR mechanisms,
and thus the security guarantees \hhf{of such TRR mechanisms} \emph{cannot} be
\hhf{properly and openly} evaluated. 

\hhf{In fact, a recent work, TRRespass~\cite{frigo2020trrespass}, \hhs{shows}
that existing proprietary in-DRAM TRR mechanisms can be circumvented via
many-sided RowHammer attacks, which aim to overflow the internal tables that TRR
uses to detect aggressor rows. As such, it is critical to develop a rigorous
methodology to understand \jk{the weaknesses of TRR mechanisms and develop more
secure alternatives.} 

\hhf{\hhs{\textbf{Our goal}} is to study} in-DRAM TRR mechanisms \hhf{so that we
can understand how they operate, assess their security, and enable fully-secure
DRAM against RowHammer.}}

%% file: sections/3_overview.tex
\vspace{-1mm}
\section{Overview of U-TRR}
\label{sec:overview}
\vspace{-1mm}

\hht{\method{} is a \hhf{new} methodology for gaining visibility into Target Row
Refresh (TRR) operations. It \jk{enables system designers and researchers to
understand} how TRR detects an aggressor row, when it refreshes the victim rows
of the aggressor row\hhf{, and how many \jk{potential} victim rows it refreshes}.}
\hht{\jk{\method{} enables users to \hhs{easily}} conduct experiments that
uncover the inner workings of the TRR mechanism \jk{in an off-the-shelf DRAM
module}.}

\begin{figure}[!h]
    \centering
    \includegraphics[width=\linewidth]{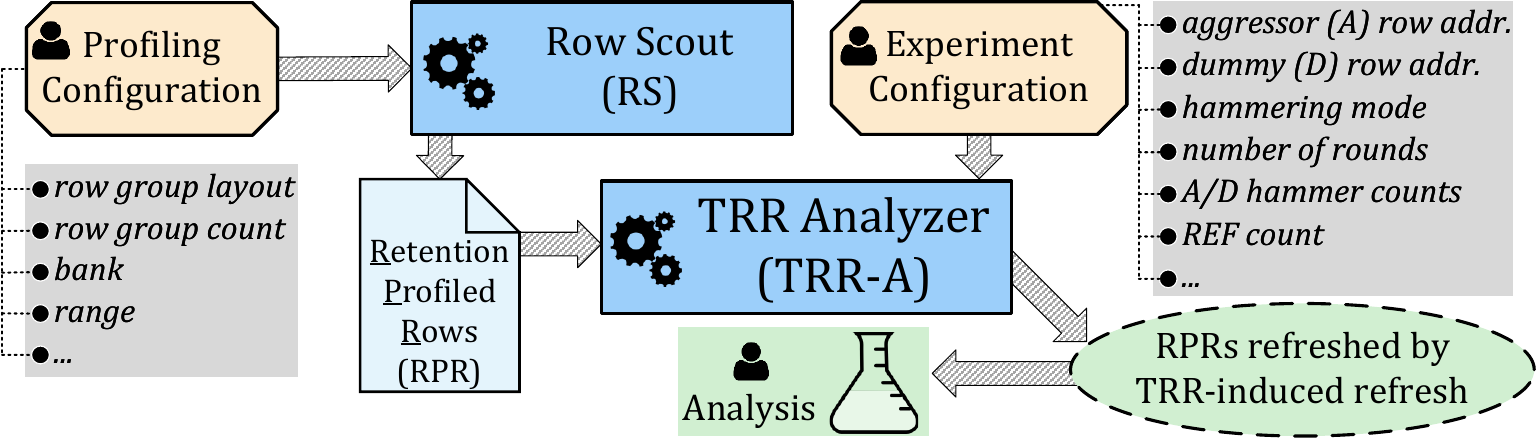}
    \vspace{-4mm}
    \caption{Overview of \method{}.}
    \vspace{-4mm}
    \label{fig:methodology_overview}
\end{figure}

Fig.~\ref{fig:methodology_overview} illustrates \hh{the two components of}
\method{}: \rtplong{} (\rtp{}) and \hhs{\trranlong{} (\trran{})}. \jk{\rtp{} \hhe{finds}} a
set of DRAM rows \hhf{that meet certain requirements as needed by \trran{}} and
\hhe{identifies} the data retention times of these rows. \jk{\trran{} uses} the
\rtp{}-provided rows \jk{to \hhs{distinguish} between TRR refreshes and regular
refreshes, and thus \hhs{builds} an understanding of the underlying TRR mechanism.} 

\subsection{Overview of \rtplong{} (\rtp{})}
\vspace{-1mm}

\hhf{The goal of \rtp{} is to identify a list of useful DRAM rows \jk{and their
retention times,} and provide this list to \trran{}. \hht{\rtp{} profiles the
retention time of a DRAM row by writing a data pattern (e.g., all
ones)~\cite{raidr,liu2013experimental} to the entire row and measuring the time
\hhf{interval for which} the row can correctly retain its data without being
refreshed.}

A useful DRAM row must satisfy two key requirements. First, the retention time
of the row should be consistent and \emph{not} vary over time based on effects
such as Variable Retention Time (VRT)~\cite{kang2014co, khan2014efficacy,
liu2013experimental, mori2005origin, qureshi2015avatar, restle1992dram,
yaney1987meta}. \jk{A consistent retention time is essential for \trran{} to
accurately infer whether or not a row has been refreshed after a specific time
interval, based on whether or not the row contains retention failures.
\hhs{\rtp{}} validates the retention time of a row one thousand times to ensure
its consistency over time. }

Second, to observe exactly which DRAM rows the TRR mechanism treats as victim
rows for each aggressor row it \hhf{detects}, \rtp{} should provide
\emph{multiple} DRAM rows
that have \hhs{the \emph{same}} retention times and that are located at certain
\emph{configurable} distances with respect to each other (\hhs{we call this} a
\emph{row group}). It is crucial to find rows with \hhs{the same} retention times in
order to observe whether or not TRR can refresh \emph{multiple} rows at the same
time. Enforcing a particular distance between rows is useful when \hht{the user
wants} to \hh{specify} an aggressor \jk{row} \hh{at a \jk{specific
\hht{distance} from} the \rtp{}-provided rows}.  These two requirements enable
reliable and precise analysis of TRR-induced refreshes by \trran{}.

As shown in Fig.~\ref{fig:methodology_overview}, the \hhs{number of row groups
(i.e., row group count)} and the relative distances of the rows \jk{within} the
group (i.e., row group layout) is specified in the \emph{profiling
configuration}. \method{} user also specifies a certain DRAM bank and row range
within the bank for the \rtp{} to search for the desired row groups.} We discuss
the \hhf{operation and} capabilities of \rtp{} in greater detail in
\S\ref{sec:ret_profiling}.

\vspace{-1mm}
\subsection{Overview of \hhs{\trranlong{} (\trran{})}}
\label{subsec:trran_overview}
\vspace{-1mm}

\begin{figure}[!h]
    \centering
    \includegraphics[width=.9\linewidth]{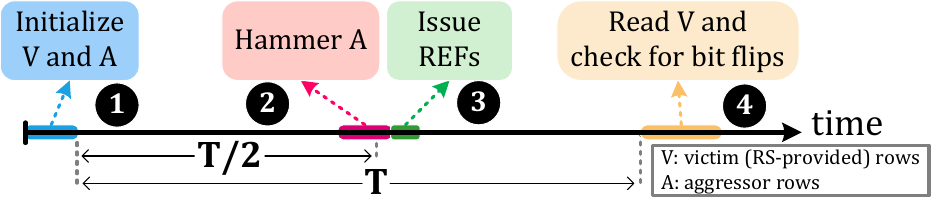}
    \vspace{-3mm}
    \caption{\hhf{General approach for detecting a TRR-induced refresh using
    \trran{}}.}
    \vspace{-3mm}
    \label{fig:general_trr_detection_approach}
\end{figure}

\hhf{The goal of \trran{} is to use \hhs{\rtp-provided rows} to determine when a
TRR mechanism refreshes a victim row by exactly distinguishing between TRR
refreshes and regular refreshes, and thus build an understanding of the
underlying TRR operation.} \trran{} \hhf{runs} \hh{a RowHammer attack and}
monitors retention failures in \hhs{\rtp{}-provided rows} to determine
\emph{when} TRR \hh{refreshes any of these rows}. \hhf{As
Fig.~\ref{fig:methodology_overview} \hhe{shows}, \trran{} \hh{operates based on
an \emph{experiment configuration}}, which includes several parameters we
discuss in \S\ref{subsec:trr_operation}.}
\hh{Fig.~\ref{fig:general_trr_detection_approach} shows} the three steps a
\trran{} experiment generally follows:
\begin{enumerate} 
    \item \trran{} \hht{uses \rtp{}-provided rows as victim rows and}
        initializes~\incircle{1} \hh{them} by writing \hhf{into them} the same data pattern
        that is used during retention profiling \hh{with \rtp{}}. \hh{Since
        \hhf{the} RowHammer \hhf{vulnerability greatly} depends on the data
        \hhf{values} stored in an aggressor
        row~\cite{kim2020revisitingrh,kim2014flipping}, \trran{} also
        initializes aggressor rows \hhf{to the data values} \hht{that the user
        specifies} in the experiment configuration.} \hhf{\trran{} waits for
        half of the victim rows' retention time \hhs{($\frac{T}{2}$)} without performing any refreshes
        or accesses.}
    \item \trran{} hammers~\incircle{2} the aggressor rows and issues
        \cmdrefresh{}~\incircle{3} \hh{commands} \hhf{based on} \hh{the
        experiment configuration}. 
    \item After \hhf{again waiting for} the half of the \hhf{victim rows'}
        retention time \hhs{($\frac{T}{2}$)}, \hh{excluding the} time spent on hammering and refresh
        \hhf{during step (2)}, \trran{} \hh{reads the} victim rows \hh{and
        compares~\incircle{4} \hhf{the data stored in them} against the} initial
        \hh{data \hhf{value} pattern written to them in} \hhf{step (1)}. \hh{A
        victim row with no} \hht{bit flips} \hh{indicates} that either a
        TRR-induced or a regular refresh operation targeted the victim row
        \hht{while serving the \cmdrefresh{} commands} in \hhf{step (2)}.
        \hhf{\trran{} easily distinguishes between a TRR-induced and \hhs{a}
        regular refresh as the latter refreshes a row periodically (e.g.,
        \jk{once in every} $8K$ \cmdrefresh{} commands).} \hhs{The user can then
        examine the TRR-induced refresh patterns to gain insight into the
        underlying TRR implementation.}
\end{enumerate}

\vspace{-3mm}

\hht{\trran{} offers flexibility in experiments via parameters in the
profiling and experiment configurations, which enable analyzing different TRR
implementations with low effort.} \hh{\S\ref{sec:analyzing_refs} \hhn{provides details}.}

\subsection{\hhf{Required Experimental Setup for \method{}}}
\label{subsec:softmc}
\vspace{-1mm}

\begin{figure}[!h]
    \centering
    \includegraphics[width=.8\linewidth]{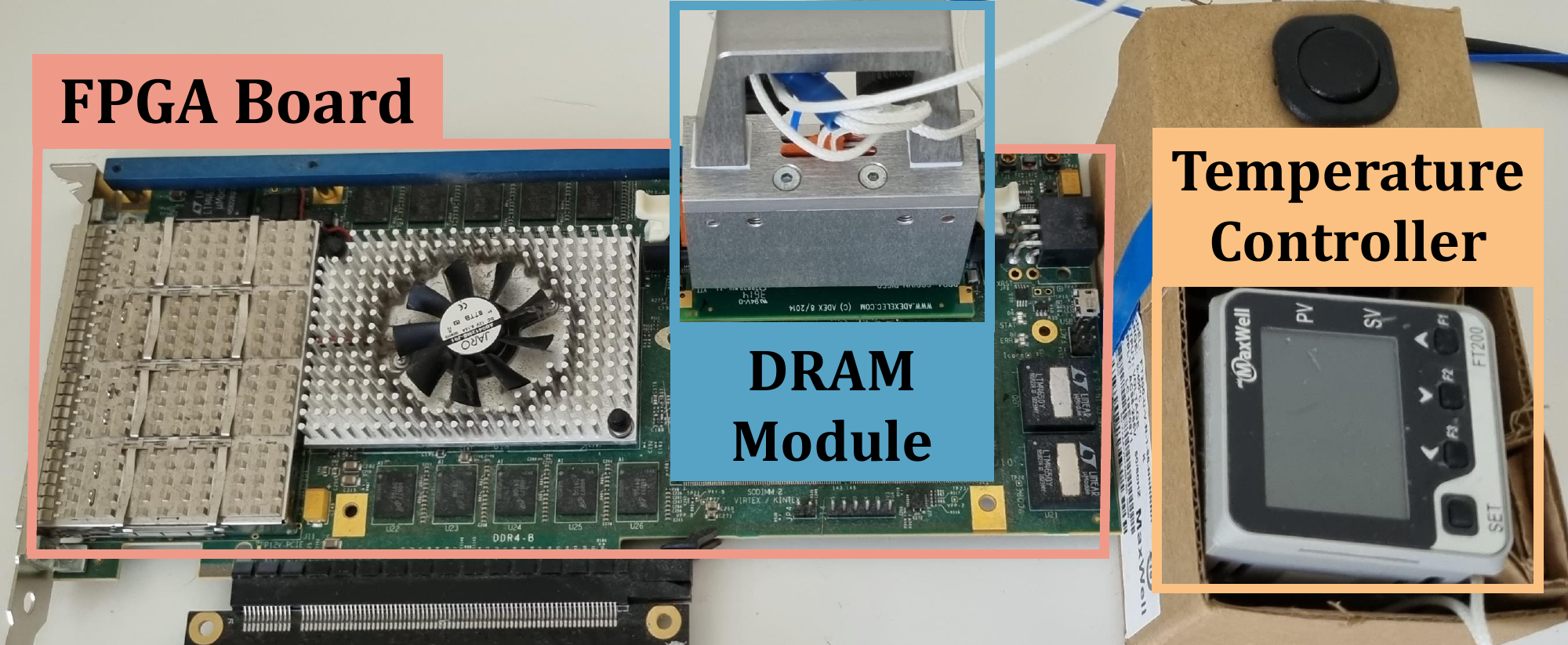}
    \vspace{-3mm}
    \caption{DDR4 \hhf{SO-DIMM} SoftMC~\cite{hassan2017softmc} experimental setup.}
    \vspace{-3mm}
    \label{fig:softmc_photo}
\end{figure}

Both the \rtp{} and \trran{} tools require a way to directly interface with a
DRAM \hht{module} at a DDR-command level. \hhf{This is} because the tools need
to \hh{accurately} control when an individual DDR command (e.g., \cmdact{},
\cmdrefresh{}) is issued to the DRAM \hht{module}. However, existing systems
based on commodity CPUs can access DRAM \hh{\emph{only}} using load/store
instructions. Therefore, we implement \rtp{} and \trran{} using
SoftMC~\cite{hassan2017softmc,softmc-safarigithub}, an FPGA-based DRAM testing
infrastructure that provides precise control on the DDR commands issued to
\hh{a} DRAM \hht{module}. We modify SoftMC to support testing DDR4 modules\hhf{,
as also done in~\cite{kim2020revisitingrh,olgun2021quac, orosa2021deeper,
frigo2020trrespass}}. \hht{Fig.~\ref{fig:softmc_photo} shows our experimental
SoftMC setup.} Table~\ref{table:trr_summary} provides a list of the
\hhn{\numTestedDIMMs{} DDR4 DRAM} \hht{modules} we analyze in this paper.

\vspace{-3mm}

%% file: sections/4_retention_profiling.tex
\vspace{-2mm}
\section{\hhf{Row Scout (\rtp{})}}
\label{sec:ret_profiling}
\vspace{-1mm}

\hht{\method{} uses} \hhf{data} retention failures as a side channel to
determine \hhf{if and when} a row receives a (\hhf{TRR-induced} \hh{or regular})
refresh. \hhf{As such}, \hht{to know how long a row can retain its data correctly without
being refreshed,} \method{} requires a mechanism for \hht{profiling} the
retention time of a DRAM row. \hhth{We define \emph{DRAM row retention time} as
the maximum time interval for which all cells in the row can correctly retain
their data without being refreshed.}

Unlike existing DRAM retention-time profiling
techniques~\cite{patel2017reaper,liu2013experimental,
raidr,qureshi2015avatar,das2018vrl}, \hh{\method{} does \emph{not}} require a
profiler that finds the retention time of \emph{all} rows in \hht{a DRAM chip}.
Instead, \hh{\method{} needs} \hhth{to} search for a \hhf{small set} of DRAM
rows \hhf{(i.e., tens of rows depending on the experiment)} that match certain
criteria \hhf{(\S\ref{subsec:profiler_requirements}) \jk{as specified by
the \method{} user based on the desired experiment.}}

\subsection{\hhf{Row Scout (\rtp{}) Requirements}}
\label{subsec:profiler_requirements}
\vspace{-1mm}

\hht{Depending on the experiment that the \method{} user conducts, \trran{}
needs the \hhf{data retention time} of DRAM rows that match different criteria.
We identify the following general requirements for \hhf{\rtp{} to enable it to
search} for DRAM rows suitable for \trran{}.}

\textbf{\hhf{Rows with uniform} retention time.} A TRR mechanism may
\hhf{refresh} multiple \hht{victim rows}. For instance, during a single-sided
RowHammer attack, a TRR mechanism may \hht{refresh the} row on either side of
the aggressor \emph{or} both at the same time. 
\hhf{To examine whether or not TRR can refresh multiple victim rows 
at the same time (i.e., with a single \cmdrefresh{}), \rtp{} must provide
multiple rows (i.e., a \emph{row group}) that have \hhs{the \emph{same}} retention
\hhe{time}.}

\textbf{\hhf{Relative positions of profiled rows}.} \hht{The location of a
victim row depends on the location of the aggressor rows that the \method{} user
specifies for an experiment.} For example, \hhe{for} a double-sided RowHammer
attack \hht{(see Fig.~\ref{fig:rowhammer_access}b)}, \rtp{} must provide three
rows with \hhs{the same} retention \hhe{time} that are exactly one row apart
from each other. \trran{} can \hhf{then} analyze which of the three \hht{victim
rows} get refreshed by TRR when hammering the two \hht{aggressor rows} \hh{that
are placed} between the \hht{victim rows}. \hhf{We represent the relative
positions of rows in a row group (i.e., the \emph{row group layout}) using a
notation such as \texttt{R-R-R}, where `\texttt{R}' indicates a
retention-profiled row and `\texttt{-}' indicates a distance of one DRAM row.
\rtp{} must find a row group based on the row group layout that the user
specifies.}

\textbf{\hhf{Rows in specific DRAM regions}.} TRR may treat rows in
different parts of a DRAM chip differently \hhf{by operating independently at
different granularities}. For example, TRR may operate independently
\hhf{at the granularity of a DRAM bank or a region of DRAM bank}. To
\hhf{identify} the granularity at which TRR operates, \rtp{} must
find DRAM rows within a \emph{specific region} of a DRAM chip.

\hhf{
    \textbf{Rows with consistent retention time.} 
	An \rtp{}-provided row must have a consistent retention time \jk{such that
	\method{} can accurately infer the occurrence of a TRR-induced refresh
	operation based on whether the row contains retention failures after a time
	period equivalent to the row's retention time. The \hhs{main difficulty} is
	a phenomenon known as Variable Retention Time (VRT)~\cite{kang2014co,
	khan2014efficacy, liu2013experimental, mori2005origin, qureshi2015avatar,
	restle1992dram, yaney1987meta}, which causes the retention time of certain
	DRAM cells to change over time. If an \rtp{}-provided row has an
	inconsistent retention time that was initially measured to be $T$, \method{}
	will not be able to correctly infer the occurrence of a TRR-induced refresh
	operation.\footnote{
	\hhs{\method{} fails to correctly infer a TRR-induced refresh when a row
	retains its data for \hhe{significantly} longer or shorter than $T$.} If the row retains its
	data for longer than $T$, \method{} \hhs{will} \emph{always} \hhs{infer} the
	occurrence of a TRR-induced refresh operation. If the row fails too soon
	(i.e., before $\frac{T}{2}$ or during Step~1 in
	\S\ref{subsec:trran_overview}), \method{} will always observe
	retention failures, since even a TRR-induced refresh will not be able to
	prevent the bit flip (in Step~2 in \S\ref{subsec:trran_overview}).
	Consequently, \method{} will \emph{always} infer that a TRR-induced refresh
	operation was not issued to the row.}
	To ensure consistency of a row's retention time, \rtp{} validates
	the retention time of a row \emph{one thousand times} in order to rule out
	inconsistencies \hhs{(that are due to VRT)}.} 
}

\textbf{\hhf{Rows with short retention times}.} The time it takes to finish a
single \method{} experiment depends on the retention time of \hht{the rows
\rtp{} finds}. This is because even \hhf{retention-weak} DRAM rows typically
retain their data correctly for \hhf{tens \jk{or}} hundreds of
milliseconds~\cite{liu2013experimental,raidr,patel2017reaper}, whereas
\hht{other \hhf{\trran{}} operations (e.g., reading from or writing to a row,
hammering a row, performing refresh)} often take \hhe{much} less than a millisecond. Thus,
as Fig.~\ref{fig:general_trr_detection_approach} \hhe{shows}, the duration of a
\trran{} experiment is \hhf{dominated} by retention times \hhf{(T)} of the
profiled rows. \hht{To reduce the overall experiment time,} it is critical for
\rtp{} to \hhf{identify} rows \hhe{with} \hhf{short data retention times}.

\vspace{-1mm}
\subsection{Row Scout (\rtp{}) \hht{Operation}}
\label{subsec:rtp}
\vspace{-1mm}

We design and \jk{implement} \rtplong{} (\rtp{}), a DRAM retention time
profiler\hhf{, such} that \hhf{it} satisfies the requirements listed in
\S\ref{subsec:profiler_requirements}. We implement \rtp{} using a
\hht{modified} version of SoftMC~\cite{hassan2017softmc, softmc-safarigithub}
with DDR4 support \hhf{(described in \S\ref{subsec:softmc})}.

\begin{figure}[!h]
    \includegraphics[width=\linewidth]{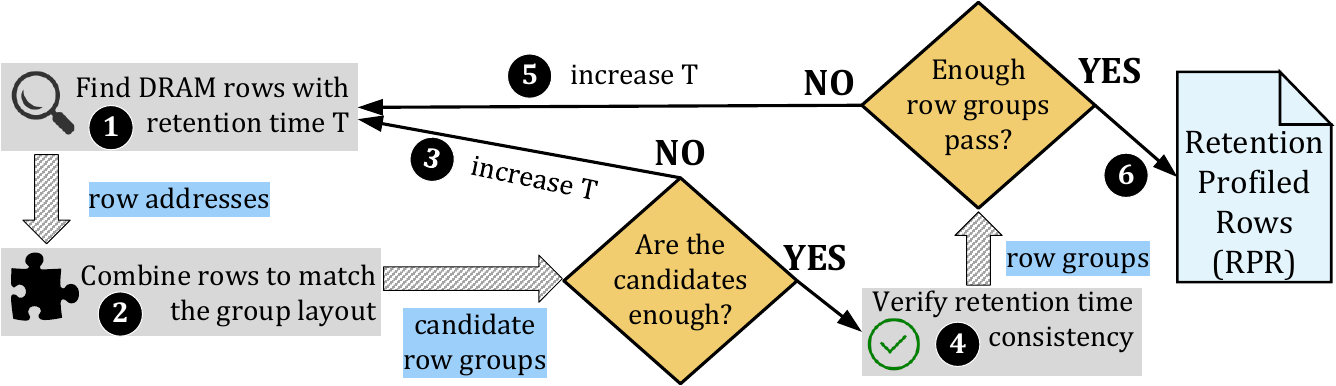}
	\vspace{-4mm}
    \caption{\hhs{Detailed} \rtp{} operation.}
	\vspace{-4mm}
    \label{fig:retprofiler_operation}
\end{figure}

\hhf{We illustrate the operation of \rtp{} in
Fig.~\ref{fig:retprofiler_operation}. \incircle{1} \rtp{} \jk{scans a full
range of DRAM rows within} a DRAM bank, as specified in the profiling
configuration (Fig.~\ref{fig:methodology_overview}), and collects the
addresses of rows that experience retention failures if not refreshed \jk{over}
the time interval $T$.  \rtp{} initially sets $T$ to a small value (e.g.,
\SI{100}{\milli\second}) in order to identify rows with small retention times
as we discuss in the requirements of \rtp{}
(\S\ref{subsec:profiler_requirements}).  \incircle{2} \rtp{} creates
candidate row groups by combining the appropriate row addresses (with retention
time $T$) that match the row group layout specified in the profiling
configuration. If the number of candidate row groups is less than the number of
row groups to find according to the profiling configuration, \incircle{3}
\rtp{} increases $T$ by a certain amount (e.g., \SI{50}{\milli\second}) and
starts over from~\incircle{1}. Otherwise, \incircle{4} \rtp{} tests each row in
a candidate row group one thousand times to ensure that all rows in the candidate
row group have a consistent retention time that is equal to $T$. If the number
of candidate row groups that pass the retention time consistency test is less
than the number of row groups to find according to the profiling configuration,
\incircle{5} \rtp{} increases $T$ by a certain amount (e.g.,
\SI{50}{\milli\second}) and starts over from~\incircle{1}. Otherwise,
\incircle{6} \rtp{} provides a list of \jk{retention time-profiled} rows to be
used by \trran{}.}

\vspace{-4mm}

%% file: sections/5_analyzing_refs.tex
\vspace{-1mm}
\section{Analyzing TRR-induced Refresh}
\label{sec:analyzing_refs}
\vspace{-1mm}

\trran{} is a configurable and extensible component in \method{} for analyzing
TRR-induced \hhf{as well as regular} refresh operations. We use \hhf{the}
\trran{} to inspect in-DRAM RowHammer mitigations in modules from the three
major DRAM vendors. We implement \trran{} on top of a modified version of
SoftMC~\cite{hassan2017softmc,softmc-safarigithub} with DDR4 support.
In \S\ref{sec:overview}, we discuss the general operation of \trran{}
using Fig.~\ref{fig:general_trr_detection_approach}. 

\vspace{-1mm}
\subsection{TRR Analyzer Requirements}
\label{subsec:trr_analyzer_reqs}
\vspace{-1mm}

We \hh{identify and} discuss four key requirements \hh{needed} to enable reverse
engineering a RowHammer mitigation mechanism. First, \hhf{to analyze the
capability of TRR in detecting multiple aggressor rows, \trran{} should allow
the user to specify one or more aggressor rows, their corresponding hammer
counts, and the order in which to hammer the aggressor rows.}

\vspace{-1mm}
\begin{req}
    \hhf{Ability to hammer multiple aggressor rows with \hhs{individually
    configurable} hammer \hhs{counts in a} configurable order}.
\end{req}
\vspace{-1mm}

The user should be able to specify dummy rows\footnote{\hhth{A dummy row
operates similarly to an aggressor row, but it can be implemented more
efficiently in a SoftMC program since a dummy row does not need to be
initialized with specific data unlike an aggressor row.}} that can be hammered
to \hhf{divert} the TRR mechanism to refresh the neighbors of a dummy row
instead of \hhf{victims of} an aggressor row.

\vspace{-1mm}
\begin{req}
Ability to specify dummy rows that are hammered in addition to the aggressor
rows.
\end{req}
\vspace{-1mm}

To \hhe{force} the TRR mechanism \hhf{to} perform an additional refresh
operation \hh{when} desired during the experiment, \trran{} should allow
flexibly issuing an \hhe{any} number of \cmdrefresh{} commands \hhf{at
arbitrary times}.

\vspace{-1mm}
\begin{req}
    Ability to flexibly issue \cmdrefresh{} commands.
\end{req}
\vspace{-1mm}

The TRR mechanism under study may retain its state beyond a single experiment,
potentially \hhf{causing the TRR mechanism to detect different rows as
aggressors depending} on previous experiments. \hh{For example, in a
\emph{counter-based TRR} \hhf{(\S\ref{subsec:vendorA_counter_based_TRR})},
the \hhf{TRR mechanism's internal} counter values updated \hht{due to} a
previous experiment \hhf{might} affect the outcome of future experiments.}
\hhs{To isolate an experiment from the past experiments}, \trran{} should reset
\hhf{TRR's internal state} to a consistent state after \hhf{each} experiment.

\vspace{-1mm}
\begin{req}
    Ability to reset \hhs{TRR mechanism's internal} state.
\end{req}
\vspace{-1mm}

\vspace{-1mm}
\subsection{TRR Analyzer \hh{Operation}}
\label{subsec:trr_operation}
\vspace{-1mm}

\begin{figure}[!h]
    \includegraphics[width=.9\linewidth]{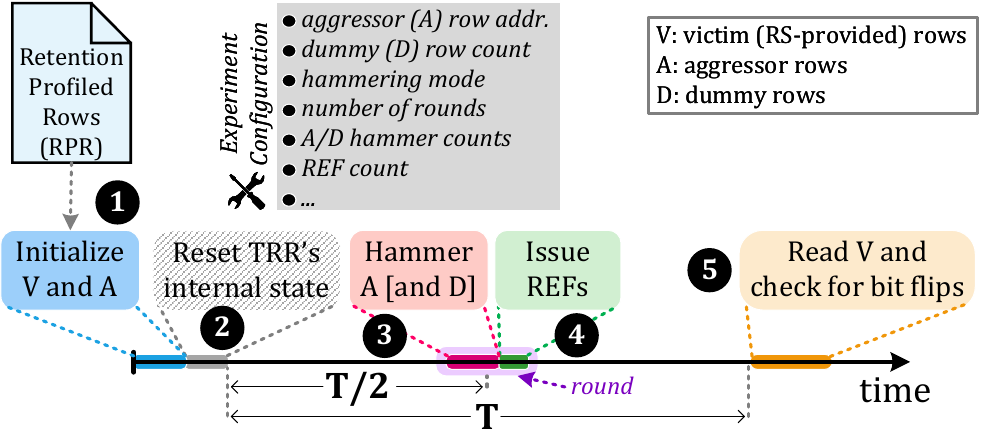}
    \vspace{-3mm}
    \caption{\hhf{\hhs{Detailed} \trran{} experiment.}}
    \vspace{-3mm}
    \label{fig:trr_analyzer_experiment}
\end{figure}

We explain how \hhs{\trranlong{} (\trran{})} satisfies all of \hhf{the
requirements described in \S\ref{subsec:trr_analyzer_reqs}} to enable detailed
experiments that \hhf{uncover} the implementation details of in-DRAM RowHammer
mitigation mechanisms. Fig.~\ref{fig:trr_analyzer_experiment} illustrates a
typical \trran{} experiment and provides a list of the experiment configuration
parameters.

\vspace{-3mm}

\hhf{    
\textbf{Initializing rows' data.} \trran{} initializes~\incircle{1} the
\rtp{}-provided rows by writing into them the same data pattern that that \rtp{}
used for profiling these rows. \trran{} also initializes the aggressor rows,
whose addresses are specified in the experiment configuration, as the RowHammer
vulnerability greatly depends on the data values stored in an aggressor
row~\cite{kim2020revisitingrh,kim2014flipping}.}

\hhf{\textbf{Resetting \hhe{internal} TRR state.} 
To reset \hhs{the TRR mechanism's} internal state~\incircle{2} \hhs{so as to
satisfy Requirement~4}, \trran{} \hhs{performs refresh for multiple refresh
periods while hammering a set of dummy rows.} \trran{} issues
\cmdrefresh{} commands at the default refresh rate (i.e., one \cmdrefresh{}
every \SI{7.8}{\micro\second}) for several (e.g., $10$) \SI{64}{\milli\second}
refresh periods. During these refresh operations, the \rtp{}-provided rows do
\emph{not} experience bit flips as they get refreshed by the regular refresh
operations. Between \hhs{two} refresh commands, \trran{} hammers a large number
(e.g., $128$) of dummy rows as many times as the DRAM timing parameters (i.e.,
\tras{} and \trp{}~\cite{micronddr4,ddr4operationhynix,kim2012case,
lee2013tiered,lee2015adaptive}) permit. \trran{} automatically selects dummy
rows from the same bank as the \rtp{}-provided and aggressor rows, since a TRR
mechanism may operate independently in each bank. To ensure that hammering the
dummy rows does \emph{not} cause RowHammer \hhs{bit flips} on the
\rtp{}-provided rows, \trran{} enforces a minimum row distance of $100$ between
a \hhs{selected} dummy row and the \rtp{}-provided rows. We find that the
operations we perform in~\incircle{2} make the TRR mechanism clear any internal
state that is relevant to rows activated in past experiments or
during~\incircle{1}. Resetting TRR's internal state is an optional step since
the user may \hhs{sometimes} want to examine how TRR operates across multiple
experiments.

 \textbf{Hammering aggressor and dummy rows.} The experiment configuration
specifies the addresses and individual hammer counts of aggressor rows that
\trran{} \hhs{accordingly} hammers~\incircle{3}. Optionally, the configuration
specifies a number of dummy rows that can be hammered to divert the TRR
mechanism away from the actual aggressor rows. \trran{} automatically
\hhs{selects} dummy rows based on the criteria for selecting dummy rows
\hhe{explained above}. The experiment configuration specifies a single hammer
count for all dummy rows; \trran{} hammers each dummy row by that count.}

\hhf{\textbf{Hammering modes.}} \hh{The order \hhe{with} which DRAM rows get hammered
can affect both the \hhe{magnitude} of the disturbance \hhe{each hammer} causes and how
the TRR mechanism \hhf{detects an aggressor row}. \trran{} supports two
\emph{hammering modes}. \hhe{\emph{Interleaved}} mode hammers each aggressor row
\hhe{one} at a time until all aggressors accumulate their corresponding hammer count.
\hhe{\emph{Cascaded}} mode repeatedly hammers one aggressor row until it
accumulates its corresponding hammer count and then does the same for the other
\hhf{aggressor rows}. In our experiments, we observe that interleaved hammering
generally causes more bit flips \hhf{(up to four orders of \hhe{magnitude})} compared
to cascaded hammering for a \hhe{given} hammer count. In \hhf{contrast}, \hhf{we
find that} cascaded hammering is more effective \hhf{at} evading the TRR
mechanism \hhf{than} interleaved hammering. Therefore, it is critical to support
\hhf{both} hammering modes.}

\hhf{\textbf{Issuing \cmdrefresh{}s.} To cause the TRR mechanism to perform
\hhs{TRR-induced} refresh operations on the victim rows, \trran{} issues a
number of \cmdrefresh{} commands~\incircle{3} according to the experiment
configuration. As shown in Fig.~\ref{fig:trr_analyzer_experiment}, \trran{}
issues the \cmdrefresh{} commands \emph{after} hammering the aggressor/dummy
rows and waiting for half of the retention time $T$ of the \rtp{}-provided rows.
This is to 1) allow TRR to potentially detect the aggressor rows hammered
in~\incircle{3} and 2) ensure that a victim row refreshed in~\incircle{4} does
\emph{not} experience retention failures until its data is read
in~\incircle{5}.}

\hhf{\textbf{Hammering rounds.}} \hht{To allow distributing hammers between
multiple \cmdrefresh{} commands,} \hh{\trran{} performs hammers in multiple
\emph{rounds}. A round consists of 1) hammering the aggressor and dummy rows and
2) issuing \cmdrefresh{} commands as the last operation of the round. \hhf{The
experiment configuration specifies the aggressor/dummy row hammer counts and the
number of \cmdrefresh{}s to issue per round.}}

\vspace{-1mm}
\subsection{Determining Physical Row Mapping}
\label{subsec:physical_row_mapping}
\vspace{-1mm}

DRAM rows that have consecutive \emph{logical} row addresses \hhf{(from the
perspective of the memory controller)} may not be physically adjacent inside a
DRAM chip~\cite{kim2012case,cojocar2020we, jung2016reverse} due to two main
reasons. First, post-manufacturing row repair \hhf{techniques
(e.g.,~\cite{kang2014co,mandelman2002challenges, nair2013archshield,
son2015cidra, horiguchi2011nanoscale, cha2017defect, ddr4, jedec2020ddr5})} may
\hhf{repair a faulty row by remapping} \hhf{the logical row address that points
to the faulty} row to a \hhf{spare row at a} different \hhf{physical} location
inside a DRAM chip. Second, a row address decoder in the DRAM chip may not
necessarily map consecutive row addresses to adjacent
wordlines~\cite{kim2012case,cojocar2020we, jung2016reverse}. The row address
decoder \hht{can} maintain the logical row address order \hhf{in physical row
space} but it may scramble \hhf{the logical-to-physical row mapping} as well
depending on the \hh{circuit implementation}.

Since a TRR mechanism should \hhs{refresh} \hhf{rows that are physically
adjacent to an aggressor row}, we need to ensure that \trran{} uses physically
adjacent rows in the experiments. For this purpose, we use two methods. First,
\hh{before we run \rtp{}, we \hh{reverse engineer the} \hhs{logical-to-physical}
row address mapping of a} DRAM chip by disabling refresh and performing
double-sided RowHammer.\footnote{\hhf{Other works~\cite{jung2016reverse,
lee-sigmetrics2017, kim2018solar, pessl2016drama, orosa2021deeper} also propose
\hhe{various} methods for reverse engineering DRAM physical layout and their methods
can also be used for our purposes.}} We analyze the rows at which RowHammer bit
flips appear\hhs{, so as} to determine the physical \hhf{adjacency of rows and
hence reconstruct the physical} row mapping. \hhs{If} the logical row address
order is preserved \hhs{in the physical space}, we simply observe RowHammer bit
flips on \hhf{logical row addresses} $R-1$ and $R+1$ \hh{as a result of}
hammering $R$. \hht{Otherwise, bit flips occur in other \hhs{logical} rows
depending on the \hhs{logical-to-physical} row address mapping of the DRAM
chip.} Second, \hhs{before} an experiment, \trran{} verifies that \hh{the given
aggressor \hhf{row can actually successfully} hammer the \hhs{\rtp{}-provided}
rows} by hammering the aggressor row a large number \hhf{of times} (i.e.,
$300K$) with refresh \hhf{disabled}. \hhf{Doing so} ensures that the
\hhs{\rtp{}-provided} or aggressor rows are \emph{not} remapped due to
post-manufacturing repair and are still \hhf{physically} adjacent to each other.

%% file: sections/6_insights.tex
\vspace{-1mm}
\section{Reverse-Engineering TRR}
\label{sec:insights}
\vspace{-1mm}

We use \method{}, the components of which we describe in
\S\ref{sec:ret_profiling} and \S\ref{sec:analyzing_refs}, to gain insights
about TRR implementations by analyzing \hh{DDR4 modules} from three major DRAM
vendors. We follow a systematic approach in our reverse engineering to gain
these insights. \hhf{First,} we discover which refresh commands perform TRR.
Second, we \hh{observe} how many \hhf{rows are concurrently refreshed by TRR}.
Third, we try to understand the strategy that \hhf{the} TRR mechanism employs
for \hh{\hhf{detecting} a DRAM row as an aggressor row \hhf{so as to}
refresh its \hhf{neighboring victim} rows}. \hht{Unless stated otherwise, we
conduct all experiments at \SI{85}{\celsius} DRAM temperature.}

Our approach leads to a number of new insights \hh{specific to} each DRAM
vendor. \hht{Table~\ref{table:trr_summary} summarizes our key findings regarding
the TRR implementations of the \hhf{\numTestedDIMMs{}} modules we test.} We
\hhf{describe} the experiments that \hh{lead us} to these insights in more
detail.

\input{tables/modules_trr_summary.tex}

\subsection{Vendor A}
\label{subsec:vendorA_trr}
\vspace{-1mm}

\hhf{Using \method{}, we find that vendor A uses two slightly different TRR
implementations in their modules as we show in Table~\ref{table:trr_summary}.}
\hhf{We explain how to use \method{} to understand the operation of the
$A_{TRR1}$ mechanism but our methodology is also applicable to $A_{TRR2}$. 
}

\vspace{-1.7mm}
\subsubsection{TRR-capable \cmdrefresh{} Commands.}

We first run an experiment to determine whether all of the \cmdrefresh{}
commands issued to DRAM can perform \hhf{TRR-induced} refresh in addition to the
regular refresh operations or only certain \cmdrefresh{} commands are
responsible for \hhf{TRR-induced refresh}. 

\hhf{To \hhf{uncover} TRR-capable \cmdrefresh{} commands, we perform experiments
that follow} the general template that we present in
Fig.~\ref{fig:trr_analyzer_experiment}. \hh{We} use \rtp{} to find $N$ row
groups that match the \texttt{R-R} layout. \hhf{Among} the profiled rows in each
row group, we designate an aggressor row, which we hammer $H$ times. We choose
$H$ so that \hhf{it does not cause RowHammer bit flips on the profiled rows} but
\hh{at the same time} \hhf{is} large enough to potentially trigger the TRR
mechanism \hh{to consider the hammered row as a potential aggressor}. To
\hhf{verify} \hh{that} $H$ hammers do not cause RowHammer bit flips, we simply
run \hhf{a separate} experiment where we 1) initialize the profiled rows, 2)
immediately after \hhs{initialization,} we hammer the profiled rows $H$ times each,
without performing any refresh, and 3) read back the profiled rows and
\hhf{verify that there are} no bit flips. 

We issue only one \cmdrefresh{} command to \hhf{individually} analyze each
refresh operation. Without a TRR mechanism, we expect to see retention failures
in \emph{all} of the profiled rows \hhf{in} \jk{\emph{almost every}} iteration of the
experiment. This is because \jk{each row is not refreshed for a long
enough period of time (i.e., for $T$ as in
Fig.~\ref{fig:general_trr_detection_approach}) such that retention failures
occur. Retention failures may not be observed during the \emph{very few}
iterations that a regular refresh operation refreshes one or more of the
profiled rows.}
\hhf{Since regular refreshes happen periodically (i.e., a row is refreshes by a
regular refresh at a fixed \cmdrefresh{} command interval
(\S\ref{subsec:regular_refresh})), \method{} easily determines when a row
is refreshed by a regular refresh.} \hh{When we observe} a profiled row \hh{with
no bit flips} \hhf{when regular refresh is \emph{not} expected}, we attribute
that to a TRR-induced refresh operation.

When we run the experiment in Fig.~\ref{fig:trr_analyzer_experiment} with $N
\geq 16$ and $H=5K$, we find an interesting pattern where we see a row group
with no bit flips \emph{only} \hhf{in} every $9^{th}$ iteration of the
experiment, i.e., \hhf{for} every $9^{th}$ \cmdrefresh{} command issued
\hh{consecutively}. This shows that, for this particular TRR design, not all
\cmdrefresh{} commands perform a \hhf{TRR-induced} refresh but only \hhf{every
$9^{th}$} of them have this capability.

\vspace{-1.5mm}
\begin{obsA}
    \hhf{Every $9^{th}$ \cmdrefresh{} command performs a TRR-induced refresh}.
\end{obsA}
\vspace{-1.5mm}

\hhf{We also find with this} experiment that, when TRR detects an aggressor row,
it simultaneously refreshes \emph{both} victim rows on each side of the detected
aggressor row with a single \cmdrefresh{}. To check if the refreshes are limited
to these two rows, we repeat the experiment using three profiled rows on each
side of the row that we hammer (i.e., we use row group layout \texttt{RRR-RRR}).
\hhe{We} observe that the TRR mechanism refreshes \emph{four} of the victim rows
closest to the detected aggressor row, i.e., two victims on each side of the
aggressor. This is likely done to protect against the probability that
\hhf{RowHammer} bit flips \hhf{can} occur in victim rows that are two rows apart
from the aggressor rows\hhf{, as demonstrated by prior
works}~\hht{\cite{yaglikci2021blockhammer,
kim2020revisitingrh,kim2014flipping, yauglikcci2021security}}. 

\vspace{-1.5mm}
\begin{obsA}
    TRR refreshes four rows that are \hhf{physically} \hh{closest} to the
    detected aggressor row. When row address $A$ is detected as an aggressor,
    TRR refreshes rows \hhf{$A\mp1$ and $A\mp2$}.
\end{obsA}
\vspace{-1.5mm}

\hhf{We next} perform a slightly different experiment to \hhf{understand in what
sequence TRR detects the hammered rows as aggressor rows in consecutive
TRR-capable \cmdrefresh{} commands}. We use two \texttt{R-R} row groups and
hammer the \hhf{aggressor rows} $H_0$ and $H_1$ times, where $H_0 << H_1$ (e.g.,
typical values we use are $H_0=50$ and $H_1=5K$). This experiment \hhf{uncovers}
that there are two different types of TRR-induced refresh operations that
alternate on every $9^{th}$ \cmdrefresh{}. \hhf{These two TRR-induced refresh
operations differ in how they detect} an aggressor row to refresh its neighbors.
The first type ($TREF_a$) always \hhf{detects} the row that has \hh{accumulated
the most hammers since \hhf{the time TRR previously detected the same row}
(e.g., the row that we hammer $H_1$ times in this experiment)}. \hh{This
\hhf{suggests} that this particular TRR mechanism might \hhf{use} a counter
table to keep track of activation counts of the accessed DRAM rows.} \hhf{The}
second type ($TREF_b$) \hhf{detects} the same row periodically every $16^{th}$
instance of $TREF_b$. \hh{We anticipate that $TREF_b$ uses a pointer that refers
to an entry in the counter table \hhf{that has $16$ entries}. $TREF_b$ refreshes
the \hht{neighbor rows of the} row address associated with the table entry that
the pointer refers to. After performing a \hhf{TRR-induced refresh}, $TREF_b$
increments the pointer to \hhe{refer} to the next entry in the table. This is
our hypothesis \hhf{as} to why $TREF_b$ repeatedly \hhe{detects} the same row
once every 16 instances of \hhf{$TREF_b$}, and \hhf{detects} other activated
rows that are in the counter table during other instances \hhf{of $TREF_b$}.} In
\S\ref{subsec:vendorA_counter_based_TRR}, we \hhf{uncover the exact}
\hhs{reason} why we see the neighbors of the same row refreshed every $16^{th}$
$TREF_b$ operation.

\vspace{-1.5mm}
\begin{obsA}
    The TRR mechanism performs two types of TRR-induced refresh operations
    ($TREF_a$ and $TREF_b$) \hh{that both use a 16-entry counter table} \hhs{to
    detect aggressor rows}.\\
    \hhf{$TREF_a$: \hhf{Detects} the row that corresponds to the table entry with the highest counter value.\\
    $TREF_b$: Traverses the counter table by \hhf{detecting} a row that corresponds to one table entry at each of its instances.
    }
    
    \label{obsA:trr_types}
\end{obsA}
\vspace{-1.5mm}

\vspace{-2mm}
\subsubsection{Counter-based TRR.}
\label{subsec:vendorA_counter_based_TRR}

\hhf{Observation}~\ref{obsA:trr_types} \hhf{indicates} that the TRR mechanism is
capable of determining which \hhf{single} row is activated (i.e., hammered) more
than the others. This \hhf{suggests} that the TRR mechanism implements a set of
counters it associates with the accessed rows and increments the corresponding
counter upon a DRAM row activation. We perform a set of experiments using the
\method{} methodology to understand more \hh{about} this counter-based TRR
implementation \hhf{we hypothesize about}.

To find the maximum number of rows that the TRR mechanism can keep track of, we
perform an experiment where we use $N$ \texttt{R-R} row groups and hammer the
rows between the profiled rows $H$ times each in \hhf{cascaded hammering mode
(\S\ref{subsec:trr_operation})}. \hhe{We} use $H=1K$ and vary $N$ in \hh{the}
range \hhf{$1 \leq N \leq 32$}. When we \hhf{repeatedly} run the experiment, we
observe that all profiled rows are eventually refreshed by $TREF_a$ or $TREF_b$
when $1 \leq N \leq 16$. However, with $N \geq 16$, we start observing profiled
rows that are never refreshed (except when they are refreshed due to regular
refresh operations as we discuss in \S\ref{subsec:regular_refresh}). Thus, we
\hhf{infer} that \hhs{this particular TRR mechanism has a counter table}
capacity for 16 different row addresses. \hhf{We also} observe that the $TREF_b$
operations \hhf{detect} rows by repeatedly iterating over the counter table
entries, such that a $TREF_b$ \hhf{detects} a row associated with one entry and
the \hhf{next} $TREF_b$ \hhf{detects} the row associated with the next entry. We
find this to be the reason for why every $16^{th}$ $TREF_b$ \hhf{detects} the
same aggressor row \hhf{when the same set of aggressor rows are repeatedly
hammered}. \hhf{Using row \hhs{groups} from different banks, we uncover that the
TRR mechanism keeps track of $16$ different rows in \emph{each} bank, suggesting
that each bank implements a separate counter table.}

\vspace{-1.5mm}
\begin{obsA}
    The TRR mechanism counts how many times DRAM rows are activated using a
    \hhs{per-bank} counter table, which can keep track of \hhf{activation
    counts} to $16$ different row addresses.
\end{obsA}
\vspace{-1.5mm}

\hhf{We next} try to find how TRR decides which row to evict from the counter
table when a new row is to be inserted. We perform an experiment where we check
if the TRR mechanism evicts the entry with the smallest counter value from the
counter table. In the experiment, we use 17 \texttt{R-R} row groups and hammer
\hhf{the row between the two retention-profiled rows} in each group. We hammer
the aggressors in the following order. First, we hammer one of the aggressors
$H_0$ times. Next, we hammer the remaining 16 aggressors $H_1$ times, where $H_0
< H_1$ \hhf{(e.g., $H_0 = 50$ and $H_1 = 100$)}. \hhf{Even after running} the
experiment for \hhf{thousands of} iterations, we observe that \hhf{the TRR
mechanism never identifies the row that is hammered $H_0$ times as an aggressor
row}. This \hhf{indicates} that the counter table entry with the smallest
counter value is evicted from the table upon inserting a new row address (i.e.,
the last row that we hammer $H_1$ times in this case) into the table. 

\vspace{-1.5mm}
\begin{obsA}
    When inserting a new row into the counter table, TRR evicts
    the row with the smallest counter value.
\end{obsA}
\vspace{-1.5mm}

We have already observed that a DRAM row activation increments the corresponding
counter in the table. However, we do not \hh{yet} know \hhf{whether or not} a
TRR-induced refresh operation updates the corresponding counter value of the
\hhf{detected} aggressor row \hhs{(e.g., resets the counter to 0)}. To check if
a TRR-induced refresh updates the corresponding counter, we \hhf{conduct
another} experiment with two \texttt{R-R} row groups, where we hammer the two
aggressors $H_0$ and $H_1$ times. When we run the experiment multiple times with
$H_0 < H_1$, we notice that $TREF_a$ \hhf{detects} an aggressor row based on how
many hammers the \hhf{aggressor} row accumulated since the last time it is
\hhf{detected} by $TREF_a$ or $TREF_b$. For example, with $H_0=2K$ and
\hht{$H_1=3K$}, the corresponding counters accumulate $36K$ and $54K$ hammers,
respectively, assuming the $18^{th}$ \cmdrefresh{} performs
$TREF_a$.\footnote{\hhf{Since $TREF_a$ and $TREF_b$ happen every $9^{th}$
\cmdrefresh{} in \hhe{an} \hhs{interleaved} manner, $TREF_a$ \hhs{happens every}
$18^{th}$ \cmdrefresh{}.}} Thus, $TREF_a$ \hhf{detects} the \hhf{aggressor} row
\hhf{that is} hammered $54K$ times and resets the corresponding counter. Until
the subsequent $TREF_a$ operation, the two counters reach $72K$ and $54K$
hammers, respectively, and $TREF_a$ \hhf{detects} the first aggressor \hhf{row}
as its counter value is higher than \hhf{that of the second aggressor row}
\hht{since} \hhf{the latter} counter \hhf{was reset earlier}. This experiment
shows that a TRR-induced refresh operation resets the counter that corresponds
to the \hhf{aggressor} row \hhf{detected} to refresh \hhf{the neighbors of}.

\vspace{-1.5mm}
\begin{obsA}
    \hht{When TRR \hhf{detects an aggressor row}, TRR resets the counter
     corresponding to the \hhf{detected} row to zero.}
\end{obsA}
\vspace{-1.5mm}

\hhf{We next} question whether, once inserted, a row address remains
indefinitely in the counter table or TRR periodically clears out
the counter table. To answer this question, we run \emph{\hh{only} once} an
experiment with one \texttt{R-R} row group and hammer the row between the
profiled rows several times to insert the aggressor rows into the counter table.
Then, we repeat the experiment many times \hhf{\emph{without} hammering the
aggressor row}. After running these experiments, we observe that the aggressor
row is \hhf{detected} by a $TREF_a$ operation only once. This is expected since
we do not access the row except in the first experiment, and once reset by the
first $TREF_a$, the corresponding counter value remains reset and never becomes
a target for $TREF_a$ again. However, we \hhf{observe} that every $16^{th}$
$TREF_b$ \hhf{detects} the \hhf{same} aggressor row and refreshes its neighbors.
We keep observing the same even after repeating the experiment $32K$ times
(i.e., issuing $32K$ \cmdrefresh{} commands that equal the number of refreshes
issued within four \SI{64}{\milli\second} nominal refresh periods). This shows that the
aggressor row remains in the counter table and keeps getting periodically
\hhf{detected} by $TREF_b$ operations. The TRR mechanism does \emph{not}
\hhf{seem to} periodically clear the counter table, for example, based on time
or the number of issued \cmdrefresh{} commands.

\vspace{-1.5mm}
\begin{obsA}
	\jk{After an entry corresponding to \hhe{a row} is inserted into the
	counter table, the entry remains in the table indefinitely until it is
	evicted \hhe{due to} insertion of a different row.} 
\end{obsA}
\vspace{-1.5mm}

\vspace{-1.5mm}
\subsubsection{Analyzing Regular Refresh.}
\label{subsec:regular_refresh}

To refresh every DRAM cell at the default \SI{64}{\milli\second} period, \hhf{the memory
controller \hhf{issues} a \cmdrefresh{} command once every
\SI{7.8}{\micro\second} according to the DDR4 specification~\cite{ddr4,micronddr4, ddr4operationhynix}.}
\hhf{In total, the memory controller issues $\approx8K$
(\SI{64}{\milli\second}$/$\SI{7.8}{\micro\second}) \cmdrefresh{} commands every
\SI{64}{\milli\second}. Therefore, it is expected that $\approx8K$ \cmdrefresh{}
commands refresh each row in the DRAM chip once to prevent a row from leaking
charge for more than \SI{64}{\milli\second}.}
\hhf{In our experiments, we observe} that the DRAM chips of vendor A internally
refresh more rows with each \cmdrefresh{} such that a row receives a regular
refresh once every $3758$ \hhf{(instead of  $\approx8K$)} \cmdrefresh{}
commands. \hhs{Thus, the DRAM chip internally refreshes its rows} with a period
\hh{even} smaller than \SI{32}{\milli\second} instead of \hh{the
\hhe{specified}} \SI{64}{\milli\second}. We suspect this \hhs{could be} an
additional measure vendor A takes \hhe{1)} to protect against
RowHammer~\cite{kim2014flipping} or \hhs{2)} in response to the decreasing
retention time as DRAM technology node \hhf{size} \hhe{becomes
smaller}~\cite{raidr,chang2014improving,mutlu2013memory,mutlu2014research}.

\begin{obsA}
    \hhe{Periodic DRAM refresh leads to internally refreshing the DRAM chip with
    a period smaller than half of the specified \SI{64}{\milli\second} refresh
    period.}
\end{obsA}
\vspace{-1.5mm}

In \S\ref{sec:case_studies}, we exploit the insights we present in this
section to craft a new DRAM access pattern that \hhf{effectively} circumvents
the protection \hhs{of} the TRR mechanism. This new \hhs{custom} access pattern
\hhf{induces} a significantly higher number of RowHammer bit flips than the
\hh{state-of-the-art access patterns presented in}~\cite{frigo2020trrespass}.

\subsection{Vendor B}
\label{subsec:vendorB_trr}
\vspace{-1mm}

\hhf{Using \method{}, we find that vendor B uses three slightly different TRR
implementations in their modules, as Table~\ref{table:trr_summary} \hhe{shows}.}
\hhf{We explain how to use \method{} to understand the operation of the
$B_{TRR1}$ mechanism. \hhs{Our} methodology is also applicable to $B_{TRR2}$ and
$B_{TRR3}$.}

\vspace{-1.7mm}
\subsubsection{TRR-capable \cmdrefresh{} commands.}

Similar to vendor A, we again start with \hhf{uncovering} which \cmdrefresh{}
commands can perform \hhf{TRR-induced refresh}. When we repeatedly run the
experiment with one or more row groups (i.e., $N \geq 16$) \hhf{while hammering
each aggressor row $5K$ times in each iteration of the experiment}, we observe
that not all \cmdrefresh{} commands perform \hhf{TRR-induced refresh}. Instead, 
we find that, in $B_{TRR1}$ \hhf{only} every $4^{th}$ \cmdrefresh{} command is
used for \hhf{TRR-induced refresh}. \hhf{Similar experiments on modules that
implement $B_{TRR2}$ and $B_{TRR3}$ uncover that every $9^{th}$ and $2^{nd}$
\cmdrefresh{} command, respectively, is used for \hhf{TRR-induced refresh}.}

\vspace{-1.5mm}
\begin{obsB}
    \hhf{Every $4^{th}$, $9^{th}$, and $2^{nd}$ \cmdrefresh{} command performs a
    TRR-induced refresh in \hhs{the three TRR mechanisms of vendor B.}}
\end{obsB}
\vspace{-1.5mm}

From the same experiment, we also observe that a TRR-induced refresh operation
refreshes only the two \hhf{neighboring rows that are immediately} adjacent to
the hammered row as opposed to vendor A's TRR implementation, which refreshes
the four \hhf{physically} closest rows to the hammered row.

\vspace{-1.5mm}
\begin{obsB}
    \hhf{The TRR mechanism refreshes the two rows physically closest to the
    detected aggressor row. When row address $A$ is detected as an aggressor,
    TRR refreshes rows $A\mp1$.}
\end{obsB}
\vspace{-1.5mm}

\vspace{-2mm}
\subsubsection{Sampling-based TRR.}

We perform experiments to show how the TRR mechanism detects the potential
aggressor rows. When we perform the experiments \hhf{that we \hhs{use} for
vendor A's modules \hhs{(described in
\S\ref{subsec:vendorA_counter_based_TRR})}}, we do \emph{not} observe
obvious patterns in the rows \hhf{detected} by TRR \hhf{\hhs{so as to} indicate
a counter-based TRR implementation}. Instead, we observe that the aggressor row
that is last hammered before a \cmdrefresh{} command is more likely to be
\hhf{detected}. In particular, when we hammer two aggressor rows $H_0=5K$ and
$H_1=3K$ times, respectively, we find that the $4^{th}$ \cmdrefresh{}
\emph{always} refreshes the neighbors of the second aggressor \hhf{row}, which
we hammer $2K$ times less than the first aggressor row. We perform experiments
with different $H_0$ and $H_1$ values and find that, when we hammer the
\hhf{second} row at least $2K$ times and issue a \cmdrefresh{}, the TRR
mechanism \hhf{consistently refreshes} the neighbors of the second row on every
$4^{th}$ \cmdrefresh{}. However, as we reduce $H_1$, the first aggressor
\hhf{row} gets \hhf{detected} by TRR with an increasing \hh{probability}. 

With further analysis, we determine that \hhe{$B_{TRR1}$} operates by sampling
the row addresses provided along with \cmdact{} commands. This sampling of
\cmdact{} commands happens with a certain probability such that $2K$ consecutive
activations to a particular row \hhf{consistently causes} the row \hhf{to be}
\hhf{detected for TRR-induced refresh}. We did not analyze this aspect of TRR
further; we suspect \hhf{(based on some \hhe{experiments})} that the sampling
does not happen truly randomly but \hhf{is} likely based on \hhe{pseudo-random
sampling of} an incoming \cmdact{}.

\vspace{-1.5mm}
\begin{obsB}
    TRR probabilistically detects aggressor rows by sampling
    row \hhe{addresses of \cmdact{} commands}.
\end{obsB}
\vspace{-1.5mm}

To determine \hhf{how many rows the TRR mechanism can sample and refresh at the
same time}, we repeat the previous experiment with the same $H_0=5K$ and
$H_1=3K$ \hht{hammer} counts but by issuing $M$ \cmdrefresh{} commands, instead
of just one, after performing the hammers. Even when we \hhf{use a large} $M$
(e.g., to 100) such that it contains multiple TRR-capable \cmdrefresh{} commands
(e.g., $25$), we \emph{never} see the neighbors of the first aggressor row to be
refreshed but \emph{always} the neighbors of the \emph{second} aggressor row.
This \hhf{suggests} that, a newly-sampled row overwrites the
\hhf{\hhe{previously-sampled} one. Therefore, we conclude that the TRR mechanism
has a capacity to sample} \emph{only} one row address. Further, we find that
\hhf{this sampling capacity is shared across \emph{all banks} in \hhs{a DRAM
chip that implements} $B_{TRR1}$ and $B_{TRR2}$}. \hhf{When} TRR samples row
$R_1$ from DRAM bank $B_1$, it overwrites a previously-sampled row $R_2$ from
bank $B_0$ \hhf{even though $R_2$'s neighbors may not have been refreshed yet}.

\vspace{-1.5mm}
\begin{obsB}
    The TRR mechanism has a sampling capacity of \hh{only} one row that is
    shared across all banks in a DRAM chip \hhf{(except for $B_{TRR3}$)}.
\end{obsB}
\vspace{-1.5mm}

Our experiments also \hhs{uncover} that a previously-sampled row address is
\emph{not} cleared when the TRR mechanism performs a \hhf{TRR-induced} refresh
on the neighbors of this aggressor row. Instead, \hhe{when a new TRR-enabled
\cmdrefresh{} is issued}, TRR refreshes \hhf{(again)} the neighbors of the same
row.

\vspace{-1.5mm}
\begin{obsB}
    A TRR-induced refresh does not clear the sampled row, and therefore the same
    row keeps \hhf{getting detected} until TRR samples another \hhf{aggressor row}.
\end{obsB}
\vspace{-1.5mm}

\subsection{Vendor C}
\label{subsec:vendorC_trr}
\vspace{-1mm}

\hhf{Using \method{}, we find that vendor C uses three slightly different TRR
implementations in their modules\hhs{, as Table~\ref{table:trr_summary} shows}.
For brevity, we \hhe{omit} the details of the experiments as they are largely
similar to the experiments for the modules of vendors A and B
(\hhs{\S}\ref{subsec:vendorA_trr} and \S\ref{subsec:vendorB_trr}). Instead, we
only \hhf{describe} our key observations.}

\hhf{We} start with running experiments to find which \cmdrefresh{} commands are
TRR-capable. Different from the \hhf{modules of \hhs{vendors} A and B}, we find
that \hhf{vendor C's modules implement} a TRR mechanism that \emph{can} perform
a \hhf{TRR-induced} refresh during the execution of \emph{any} \cmdrefresh{}
command. The TRR mechanism performs a \hhf{TRR-induced} refresh once every $17$
consecutive \cmdrefresh{} commands during \hhf{a likely} RowHammer attack. 
When \hhf{likely} not under an attack (i.e., when a \hhf{small number of} row
activations happen), TRR can defer a \hhf{TRR-induced} refresh to any of the
subsequent \cmdrefresh{} commands until it detects an aggressor row. \hhf{We do
not observe TRR-induced refresh more frequently than once in every $17$
\cmdrefresh{} commands. For $C_{TRR2}$ and $C_{TRR3}$, we find that every
$9^{th}$ and $8^{th}$ \cmdrefresh{} command, respectively, performs a
TRR-induced refresh.}

\vspace{-1.5mm}
\begin{obsC}
    \hhf{Every $17^{th}$, $9^{th}$, and $8^{th}$ \cmdrefresh{} command normally
    performs a TRR-induced refresh in \hhs{the three TRR mechanisms of vendor
    C}. A TRR-induced refresh can be deferred to a later \cmdrefresh{} if no
    \hh{potential} aggressor row is detected.}
\end{obsC}
\vspace{-1.5mm}

To \hhf{uncover} the logic behind \hhf{how} a potential aggressor row \hhf{is
detected}, we run experiments similar to \hhf{those we use} for the modules from
vendors A and B. \jk{We find that vendor C's TRR mechanism \hhf{detects}
aggressor rows only from the set of rows targeted by the first $2K$ \cmdact{}
commands (per bank) following a TRR-induced refresh operation.} 
We also find that \hhs{1)} TRR probabilistically \hhf{detects} one of the rows
activated within the first $2K$ \cmdact{} commands and \hhs{2)} the rows that
are activated earlier have a higher chance to be targeted by \hh{the subsequent}
TRR-induced refresh operation. \hhf{Discovering that TRR detects aggressor rows
based on only the first $2K$ \cmdact{} commands helped us to craft an effective
access pattern (\S\ref{subsec:access_patterns}); thus we did not further analyze
vendor C \hhe{modules} to uncover the maximum number of potential aggressor rows
TRR \hhe{tracks}.}

\vspace{-1.5mm}
\begin{obsC}
    TRR \hhe{detects} an aggressor row only among the first $2K$
    \cmdact{}\footnote{Except for the modules that implement $C_{TRR3}$
    (Table~\ref{table:trr_summary}). $C_{TRR3}$ detects an aggressor row only
    among the first $1K$ activations to each bank.} \hhf{(to each bank)}
    following a TRR-induced refresh. \hhs{Rows} \hhf{activated} earlier \hhe{are
    more likely} to be \hhf{detected} by TRR.
\end{obsC}
\vspace{-1.5mm}

Our experiments \hhf{uncover} a unique DRAM row organization in \hhf{modules
$C0$-$8$}. It appears that two consecutively addressed rows (i.e.,
\hhf{\hhs{physical} row addresses} $R$ and $R+1$ where $R$ is an even row
address) are isolated \jk{in pairs such that hammering one row (e.g., $R$) can
induce RowHammer bit flips \emph{only} in its \emph{pair row} (e.g., $R+1$), and
not \hhs{in} any other row in the bank. As expected, we also observe that TRR
issues refresh operations \emph{only} to the pair row of each aggressor row that
it identifies.} 

\vspace{-1.5mm}
\begin{obsC}
	\jk{Given any two rows, $R$ and $R+1$, where $R$ is an even number,}
	\hhe{TRR refreshes \emph{only} one of the rows (e.g., $R$) upon detecting
	the other (e.g., $R+1$) as an aggressor row.}
\end{obsC}
\vspace{-1.5mm}

%% file: tables/modules_trr_summary.tex
\begin{table*}[!ht]
    
    \centering
    \caption{Summary of \hhf{our} key observations \hhe{and results on} TRR implementations of \hhf{\numTestedDIMMs{} DDR4 DRAM} modules.
    }
    \vspace{-4mm}
    \label{table:trr_summary}

    \resizebox{\linewidth}{!}{

            \begin{footnotesize}
                \begin{tabular}{ l c c | c c c | c | c c c c c c c c}
                    \toprule

                    \mr{3.5}{\emph{Module}} & \mr{3.5}{\emph{\makecell{Date\\(yy-ww)}}}  & \mr{3.5}{\emph{\makecell{Chip\\Density\\(Gbit)}}} & \multicolumn{3}{c}{\emph{\hhf{Organization}}} & \mr{3.5}{$HC_{first}$$\dag$} & \multicolumn{8}{c}{\emph{\hhf{Our Key TRR Observations \hhe{and Results}}}}\\

                    \cmidrule(lr){4-6}
                    \cmidrule(lr){8-15}


                    \multicolumn{3}{c|}{} & \makecell{\emph{Ranks}} & \emph{Banks} & \emph{Pins} & & \emph{Version} & \emph{\makecell{Aggressor\\Detection}} & \emph{\makecell{Aggressor\\Capacity}} & \emph{\makecell{Per-Bank\\TRR}} &
                        \emph{\makecell{TRR-to-REF\\Ratio}} & \emph{\makecell{Neighbors\\Refreshed}} & \emph{\makecell{\% Vulnerable\\DRAM Rows$\dag$}} & \emph{\makecell{Max. Bit Flips\\per Row per Hammer$\dag$}} \\

                    \midrule

                    \stripe
                    A0     & 19-50 & 8 & 1 & 16 & 8 & $16K$     & $A_{TRR1}$ & Counter-based & 16 & \cmark & $1/9$ & 4 & 73.3\% & 1.16 \\
                    A1-5   & 19-36 & 8 & 1 & 8 & 16 & $13K$-$15K$ & $A_{TRR1}$ & Counter-based & 16 & \cmark & $1/9$ & 4 & 99.2\% - 99.4\% & 2.32 - 4.73 \\
                    \stripe
                    A6-7   & 19-45 & 8 & 1 & 8 & 16 & $13K$-$15K$ & $A_{TRR1}$ & Counter-based & 16 & \cmark & $1/9$ & 4 & 99.3\% - 99.4\% & 2.12 - 3.86 \\
                    A8-9   & 20-07 & 8 & 1 & 16 & 8 & $12K$-$14K$ & $A_{TRR1}$ & Counter-based & 16 & \cmark & $1/9$ & 4 & 74.6\% - 75.0\% & 1.96 - 2.96 \\
                    \stripe
                    A10-12 & 19-51 & 8 & 1 & 16 & 8 & $12K$-$13K$ & $A_{TRR1}$ & Counter-based & 16 & \cmark & $1/9$ & 4 & 74.6\% - 75.0\% & 1.48 - 2.86 \\
                    A13-14 & 20-31 & 8 & 1 & 8 & 16 & $11K$-$14K$ & $A_{TRR2}$ & Counter-based & 16 & \cmark & $1/9$ & 2 & 94.3\% - 98.6\% & 1.53 - 2.78 \\

                    
                    \cmidrule(lr){1-15}
                    \stripe
                    B0     & 18-22 & 4 & 1 & 16 & 8 & $44K$       & $B_{TRR1}$ & Sampling-based & 1 & \xmark & $1/4$ & 2 & 99.9\% & 2.13 \\
                    B1-4   & 20-17 & 4 & 1 & 16 & 8 & $159K$-$192K$ & $B_{TRR1}$ & Sampling-based & 1 & \xmark & $1/4$ & 2 & 23.3\% - 51.2\% & 0.06 - 0.11 \\
                    \stripe
                    B5-6   & 16-48 & 4 & 1 & 16 & 8 & $44K$-$50K$   & $B_{TRR1}$ & Sampling-based & 1 & \xmark & $1/4$ & 2 & 99.9\% & 1.85 - 2.03 \\
                    B7     & 19-06 & 8 & 2 & 16 & 8 & $20K$       & $B_{TRR1}$ & Sampling-based & 1 & \xmark & $1/4$ & 2 & 99.9\% & 31.14 \\
                    \stripe
                    B8     & 18-03 & 4 & 1 & 16 & 8 & $43K$       & $B_{TRR1}$ & Sampling-based & 1 & \xmark & $1/4$ & 2 & 99.9\% & 2.57 \\
                    B9-12  & 19-48 & 8 & 1 & 16 & 8 & $42K$-$65K$   & $B_{TRR2}$ & Sampling-based & 1 & \xmark & $1/9$ & 2 & 36.3\% - 38.9\% & 16.83 - 24.26 \\
                    \stripe
                    B13-14 & 20-08 & 4 & 1 & 16 & 8 & $11K$-$14K$   & $B_{TRR3}$ & Sampling-based & 1 & \cmark & $1/2$ & 4 & 99.9\% & 16.20 - 18.12 \\

                    \cmidrule(lr){1-15}
                    C0-3   & 16-48 & 4 & 1 & 16 & x8 & $137K$-$194K$ & $C_{TRR1}$ & Mix & Unknown & \cmark & $1/17$ & 2 & 1.0\% - 23.2\% & 0.05 - 0.15 \\
                    \stripe
                    C4-6   & 17-12 & 8 & 1 & 16 & x8 & $130K$-$150K$ & $C_{TRR1}$ & Mix & Unknown & \cmark & $1/17$ & 2 & 7.8\% - 12.0\% & 0.06 - 0.08 \\
                    C7-8   & 20-31 & 8 & 1 & 8 & x16 & $40K$-$44K$ & $C_{TRR1}$ & Mix & Unknown & \cmark & $1/17$ & 2 & 39.8\% - 41.8\% & 9.66 - 14.56 \\
                    \stripe
                    C9-11  & 20-31 & 8 & 1 & 8 & x16 & $42K$-$53K$ & $C_{TRR2}$ & Mix & Unknown & \cmark & $1/9$ & 2 & 99.7\% & 9.30 - 32.04 \\
                    C12-14 & 20-46 & 16 & 1 & 8 & x16 & $6K$-$7K$ & $C_{TRR3}$ & Mix & Unknown & \cmark & $1/8$ & 2 & 99.9\% & 4.91 - 12.64 \\
                    
                    \bottomrule
                \end{tabular}
                
            \end{footnotesize}
                    
    } 
    \begin{scriptsize}
    {\raggedright 
        $\dag$We report the minimum and maximum
        $HC_{first}$, \%~\emph{Vulnerable DRAM Rows}, \hhs{and    
        \emph{Max. Bit Flips per Row per Hammer}} for table rows containing \emph{multiple}
        DRAM modules.

        $HC_{first}$: Minimum activation count per aggressor row in double-sided RowHammer to cause a bit flip. | \emph{Version}: \hhe{Unique} identifier for different TRR implementations we \hhs{observe} across DRAM vendors.
                    
        \emph{Aggressor Detection}: \hhe{Main} method used to detect an aggressor row. | \emph{Aggressor Capacity}: \hhe{Maximum} number of potential aggressor rows TRR can track.
        
        \emph{Per-Bank TRR}: Indicates whether a TRR mechanism operates independently in each bank or is shared across banks. | \emph{TRR-to-REF Ratio}: \hhe{Fraction} of TRR-capable \cmdrefresh{}s \hhn{out of} all \cmdrefresh{}s.
        
        \emph{Neighbors Refreshed}: \hhe{Number} of \hhs{neighboring victim rows} refreshed by a TRR-induced refresh. | \emph{\% Vulnerable DRAM Rows}: \hhe{Fraction} of DRAM rows vulnerable to our custom access patterns.
        
        \emph{Max. Bit Flips per Row per Hammer}: \hhe{Maximum} number of bit flips \hhe{observed} in \hhe{any victim} row per \hhe{each hammer to an aggressor row} between two \cmdrefresh{}s.\par
    }
    \end{scriptsize}

    \vspace{-5mm}
\end{table*}

%% file: sections/7_case_studies.tex
\vspace{-1mm}
\section{Bypassing TRR \hhf{Using \method{} Observations}}
\label{sec:case_studies}
\vspace{-1mm}

\method{} \hh{uncovers} critical characteristics of the TRR mechanisms different
\hh{DRAM} vendors implement in \hh{their} chips. We leverage those
\hh{characteristics} to craft \jk{custom} DRAM access patterns that hammer
\hh{an aggressor row} \jk{such that TRR \hhs{cannot} refresh the aggressor row's
neighbors (i.e., victim rows) in a timely manner.} 
\hhf{Our results show that these new custom access patterns greatly increase
RowHammer bit flips on \jk{the} \numTestedDIMMs{} DDR4 modules we \hhe{test}.}

\vspace{-1mm}
\subsection{\hhf{Custom} RowHammer Access Patterns}
\label{subsec:access_patterns}
\vspace{-1mm}

\textbf{Vendor A.} Using \method{}, we find that \hhf{vendor A's modules
implement} a counter-based TRR ($A_{TRRx}$\footnote{\hhf{We refer to
all versions of TRR mechanisms that vendor A's modules implement (i.e.,
$A_{TRR1}$ and $A_{TRR2}$) as $A_{TRRx}$. We use a similar terminology \hhs{for
other vendors}.}}), the details of which \hhe{are} in
\S\ref{subsec:vendorA_trr}. Since $A_{TRRx}$ evicts the entry with the lowest
counter value when \jk{inserting a new entry to the table}, a \hhf{custom}
RowHammer access pattern \hhf{that takes advantage of our \method{} analysis}
should first hammer \hhf{two} aggressor rows in \jk{a} double-sided manner and
then \jk{evict} the \hhf{two} aggressor rows \jk{from the table} by
hammering other rows (\jk{i.e.,} \emph{dummy rows}) \jk{within} the same bank
\hhf{during the remaining time until the memory controller issues a
\cmdrefresh{} command.}\footnote{\hhf{The memory controller issues a \cmdrefresh{}
once every \SI{7.8}{\micro\second} when using the default \SI{64}{\milli\second}
refresh period. This allows at most 149 hammers to a single DRAM bank assuming
typical activation (\SI{35}{\nano\second}), precharge (\SI{15}{\nano\second}),
and refresh (\SI{350}{\nano\second})
latencies~\cite{micronddr4,liu2013experimental,chang2014improving}.}}

We \hhf{show \hhs{how} we} can hammer two aggressor rows ($A_0$ and $A_1$) in
\jk{a} double-sided manner without \hhf{allowing} $A_{TRRx}$ \jk{to} refresh
their \hhf{victim rows}. First, the attacker should synchronize the memory
accesses with the periodic \cmdrefresh{} commands\footnote{\hht{A recent
work~\cite{deridder2021smash} shows how to detect when a memory controller
issues a periodic \cmdrefresh{} from an unprivileged process
\hhe{and} from a web browser using JavaScript.}} in order to hammer $A_0$
and $A_1$ \hhf{\emph{right after}} the memory controller issues a \cmdrefresh{}.
After hammering the two aggressor rows, the attacker \jk{should then} use the
remaining time until the next \cmdrefresh{} to hammer dummy rows in order to
\hh{steer} $A_{TRRx}$ \hh{to \jk{identify} one of the dummy rows \hhf{(and
\emph{not} rows $A_0$ and $A_1$)} as potential aggressors \hhf{\jk{and} refresh
\jk{the dummy rows'} neighboring victim rows}}. The particular access pattern
that leads to the largest number of bit flips is hammering
$A_0$ and $A_1$ 24 times each, followed by hammering 16 dummy rows 6 times each.
\hh{We discover the access pattern that maximizes the bit flip count by sweeping
the number \hhs{of hammers to} \hhf{\hhe{$A_0$ and $A_1$} and
adjusting the number of hammers to the 16 dummy rows based on the time that
remains until the next \cmdrefresh{} after hammering the aggressors}.}

\textbf{Vendor B.} \jk{$B_{TRRx}$ operates by probabilistically sampling a
single row address from all \cmdact{} (\S\ref{subsec:vendorB_trr})
commands issued \hhf{(across all banks for $B_{TRR1}$ and $B_{TRR2}$)} to DRAM.
\hhf{Rows neighboring the sampled row are refreshed during a TRR-induced
refresh operation that happens once in every $4$, $9$, and $2$ \cmdrefresh{}
commands for $B_{TRR1}$, $B_{TRR2}$, and $B_{TRR3}$, respectively.}}
\jk{\hhs{To} maximize the probability of $B_{TRRx}$ \hhs{detecting} a dummy row
instead of the aggressor row, our custom access pattern \hhs{maximizes} the
number of \hhe{hammers} to dummy rows after hammering the aggressor rows and
before every \hhf{TRR-induced refresh operation}.} 
\jk{Our custom access pattern first} hammers rows $A_0$ and $A_1$ immediately
following \hh{a} \hhf{TRR-induced} refresh. Then, it simultaneously hammers a
\hhf{single dummy row in each of four} banks\footnote{\hhf{We do not hammer a
dummy row in more than four different banks due to the Four-Activation-Window
(\tfaw{}~\cite{micronddr4, ddr4,ddr4operationhynix}) DRAM timing constraint.}}
to perform a large number of dummy row activations \hh{within the limited time}
until the next \hhf{TRR-induced} refresh.\footnote{\hhf{For $B_{TRR3}$, which
separately samples \cmdact{} commands to each bank, we hammer a dummy row from
the aggressor row's bank.}} We find that $220$ hammers per aggressor row
(leaving 156 hammers \hhf{for each dummy row in the four banks}) within a window
of four consecutive \cmdrefresh{} commands causes RowHammer bit flips even in
the least RowHammer-vulnerable module of the \numTestedDIMMsFromB{} vendor B
modules that we use in our experiments.

\textbf{Vendor C.} \hhf{$C_{TRR1}$, $C_{TRR2}$, and $C_{TRR3}$ have the ability
to perform a \hhf{TRR-induced} refresh once in every $17$, $9$, and $8$
\cmdrefresh{} commands, respectively}, and \hhf{they} can defer the
\hhf{TRR-induced} refresh to a later \cmdrefresh{} until \hhf{a potential}
aggressor row is detected \hh{(\S\ref{subsec:vendorC_trr})}. $C_{TRRx}$ does not
keep track of more than $2K$ \cmdact{} commands that follow a TRR-induced
refresh operation and rows activated earlier in the set of $2K$ \cmdact{}
commands are more likely to be \hhe{detected}. Therefore, we craft a
\hhf{custom} RowHammer access pattern that follows a TRR-induced refresh
operation with a large number (e.g., $2K$) of dummy \hhf{row} activations and
then hammers the aggressor rows $A_0$ and $A_1$ until \hhf{the next TRR-induced
refresh operation}. \hhf{To properly execute this access pattern, it is critical
to synchronize \hhs{the dummy and aggressor row hammers} with \hhs{TRR-enabled}
\cmdrefresh{} \hhs{commands}.}

\vspace{-1mm}
\subsection{\hhf{Effect on RowHammer} Bit Flip Count}
\label{subsec:overall_bit_flips}
\vspace{-1mm}

\hhf{We \hhs{implement and} evaluate the different custom DRAM access patterns
that are used to circumvent $A_{TRRx}$, $B_{TRRx}$, and $C_{TRRx}$ on \hhs{our}
FPGA-based SoftMC platform~\cite{hassan2017softmc}
\hhs{(\S\ref{subsec:softmc})}. The SoftMC program executes each custom access
pattern for a fixed interval of time (determined by each chip's TRR-induced
refresh frequency), while also issuing \cmdrefresh{} commands once every
\SI{7.8}{\micro\second} to comply with the vendor-specified default refresh
rate.}

\begin{figure}[!h]
    \centering
    \hspace{-0.2em}
    \begin{subfigure}[t]{.3\linewidth} 
        \centering
        \includegraphics[height=1.05in]{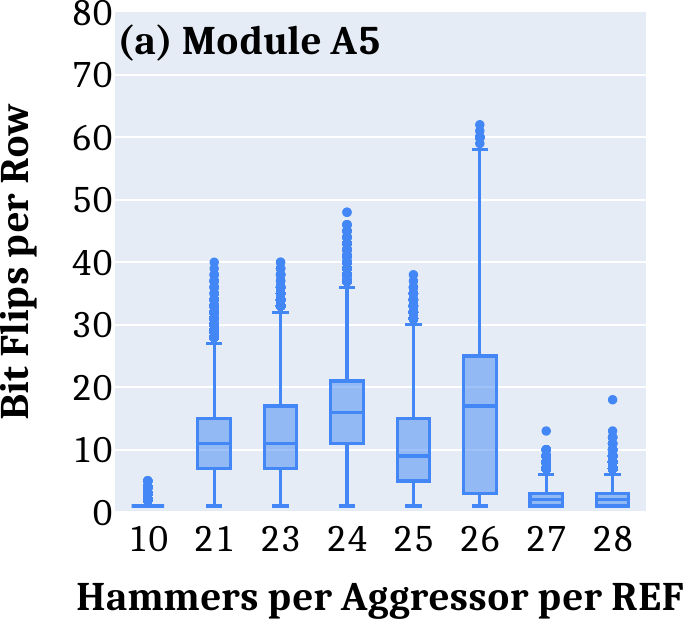}
    \end{subfigure}
    \quad
    \hspace{0.2em}
    \begin{subfigure}[t]{.28\linewidth}
        \centering
        \includegraphics[height=1.05in]{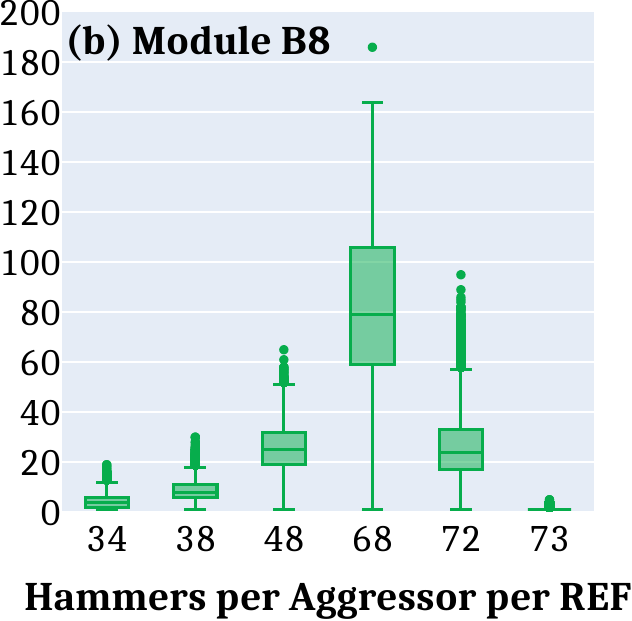}
    \end{subfigure}
    \quad
    \begin{subfigure}[t]{.3\linewidth}
        \centering
        \includegraphics[height=1.03in]{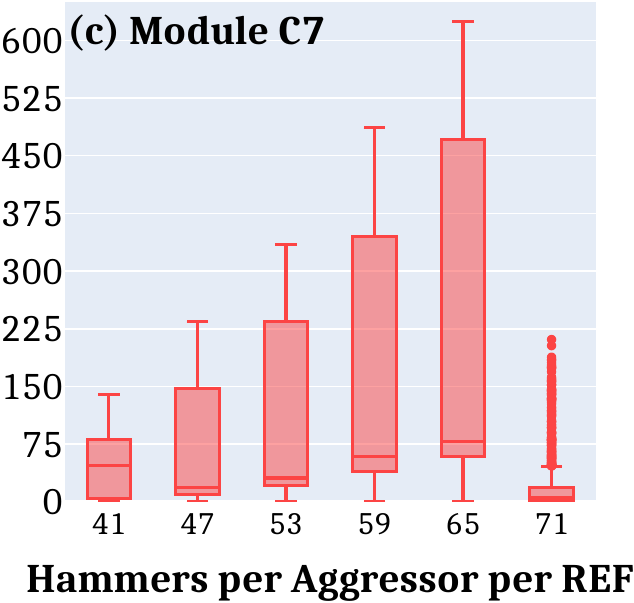}
    \end{subfigure}
    \vspace{-4mm}
    \caption{\hhs{Distribution of} \hhf{bit flips per DRAM row} for different
    aggressor hammer counts \hhs{in} \hhf{three representative modules}.}
    \label{fig:bitflips_per_bank}
    \vspace{-4mm}
\end{figure}

\hhf{Fig.~\ref{fig:bitflips_per_bank} shows the \hhs{distribution of} number
of bit flips per row as box-and-whisker plots\footnote{\hhf{The lower and upper
bounds of the box represent the first quartile (i.e., the median of the first
half of a sorted dataset) and the second quartile (i.e., the median of the
second half of a sorted dataset), respectively. The median line is within the
box. The size of the box represents the inter-quartile range (IQR). The whiskers
are placed at $1.5*IQR$ on both sides of the box. The outliers are represented
with dots.}}
in modules $A5$, $B8$, and $C7$\footnote{\hhs{We analyze $A5$, $B8$, and $C7$ as
they are the modules that experience the most RowHammer bit flips and implement
$A_{TRR1}$, $B_{TRR1}$, and $C_{TRR1}$, respectively.}} when sweeping the number
of hammers issued to aggressor rows in each custom access pattern. \hhs{The
x-axis is normalized so that it shows the average number \hhe{of} hammers
performed between two \cmdrefresh{}s to a single aggressor
row.}\footnote{\hhs{The x-axis shows the number of hammers per aggressor row per
\cmdrefresh{} to enable easy comparison of the effectiveness of different
RowHammer patterns across different modules. \hhe{Our actual experiments
perform} the aggressor and dummy row hammers as required by each custom
RowHammer access pattern \hhe{described} in \S\ref{subsec:access_patterns}.}}
\hhs{We show different hammers per aggressor per \cmdrefresh{} for each module
as the number of hammers we can fit between two \cmdrefresh{}s depend on the
\hhe{custom} RowHammer patterns we craft (\S\ref{subsec:access_patterns}).} Each
access pattern uses a fixed number of dummy rows as described in
\S\ref{subsec:access_patterns}. 
We perform the maximum number of hammers that fit \hhs{between two
\cmdrefresh{}s}. Therefore, a lower number of aggressor row \hhs{hammers}
translates to a higher number of dummy row hammers.}

\vspace{-4mm}

\textbf{Vendor A.} We observe the highest \hhf{bit flip count} (i.e.,
\hhs{up to 62 bit flips in a row}) when using \hhs{$26$} hammers per aggressor \hhf{row},
in which case each of the $16$ \hhf{dummy rows} are hammered $6$ times. The
number of bit flips decreases as we hammer the aggressors more than \hhs{$26$} times.
This is because the aggressors become less likely to be evicted from the counter
table as more activations \hhf{to them} increment the corresponding counters to
higher values. In contrast, the \hhf{aggressor rows} become more likely to be
evicted from the counter table when they are hammered \hhf{fewer} than \hhs{$26$}
times each. However, we still observe a \hhf{smaller bit flip count} with
\hhf{fewer} than \hhs{$26$} hammers per aggressor because fewer hammers are
\hhf{insufficient} \hht{for many \hhf{victim} rows} to \hhf{experience}
RowHammer bit flips.

\textbf{Vendor B.} The number of bit flips gradually increases with the number
of hammers per aggressor row. This \jk{increases to} a point where too many
aggressor row activations leave an insufficient time to perform enough dummy row
activations to ensure that a dummy row is sampled to replace an aggressor row
for the subsequent \hhf{TRR-induced} refresh. According to our experiments, at
least $12$ \hhs{total} dummy row activations \hhs{simultaneously performed in
four banks} (leaving enough time to perform \hhs{$73$ hammers per aggressor})
are needed to induce RowHammer bit flips. We observe the maximum number of bit
flips with \hhs{$68$} hammers per aggressor row.

\textbf{Vendor C.} We observe bit flips appear when a dummy row is
initially hammered a large number of times to make the subsequent aggressor row
activations less likely to be tracked by $C_{TRRx}$. The access pattern causes
bit flips when performing at least $252$ dummy hammers \hhs{(right after a
TRR-enabled \cmdrefresh{})} prior to \hhs{continuously} hammering the aggressor
rows \hhs{until the next TRR-enabled \cmdrefresh{}. This} leaves time to perform
\hhs{$71$} hammers per aggressor row \hhs{per \cmdrefresh{} \hhe{on} average}.
\hhs{We observe the maximum number of bit flips with \hhs{$65$} hammers per
aggressor row.}

\vspace{-1mm}
\subsection{\hhf{Effect on} Individual Rows}
\label{subsec:row_vulnerability}
\vspace{-1mm}

To mount a successful system-level RowHammer attack, it is critical to force the
operating system to place sensitive data in \hhe{vulnerable} rows. To
make this task easier, it is important to induce RowHammer bit flips in as many
rows as possible. \hhf{Ideally,} all rows should be vulnerable to RowHammer
from an attacker's perspective.

\begin{figure*}[tp]
    \centering
    \includegraphics[width=.85\linewidth]{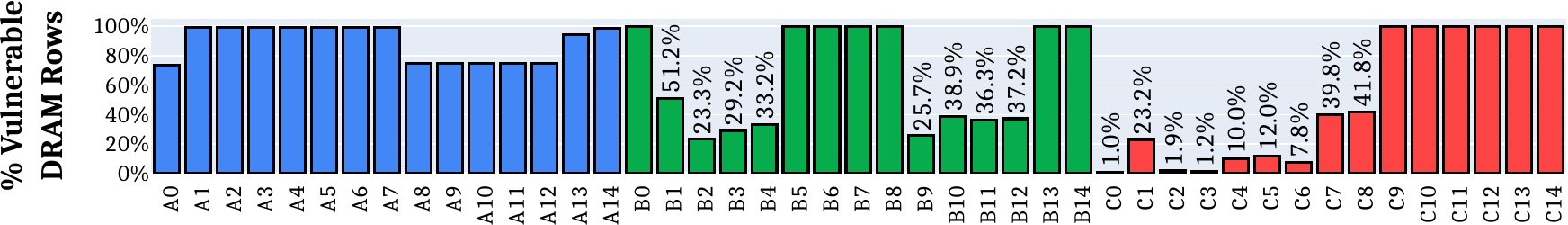}
    \vspace{-4mm}
    \caption{\hhf{Percentage of rows that experience at least one RowHammer bit
    flip using our custom RowHammer access patterns}.}
    \vspace{-5mm}
    \label{fig:vulnerable_rows}
\end{figure*}

Fig.~\ref{fig:vulnerable_rows} \hhf{shows} the percentage of vulnerable DRAM
rows, i.e., rows \hhf{that experience} at least one \hhf{RowHammer} bit flip
\hhf{with our custom RowHammer access patterns
(\S\ref{subsec:access_patterns})}, \hhf{as a fraction of} all rows in a
bank of \hhe{the tested \numTestedDIMMs{} modules}. We report data for a single
bank\footnote{\hhf{We test a single bank to reduce the experiment time. To
ensure that the results are similar across different banks, we tested multiple
banks from several modules}.} from each module. For each DRAM module, we use a
different number of hammers per aggressor that results in the highest percentage
of vulnerable rows in the corresponding module \hht{(see
\S\ref{subsec:overall_bit_flips})}.\footnote{\hhf{When using the
conventional single- and double-sided RowHammer, we do \emph{not} observe RowHammer bit
flips in any of the \numTestedDIMMs{} DDR4 modules \jk{(as expected from our
understanding of the TRR implementations \hhe{and from}~\cite{frigo2020trrespass})}.}}

In many \hhf{(i.e., $8$, $7$, \hhs{and $6$}, respectively)} modules from vendor
\hhs{A, B, and C} we see bit flips in \hhf{more than} \vulnRowsMaxPct\% of the
rows. This shows that the \hhf{custom} access \hhf{patterns} we use \hhf{are
effective at} circumventing the \jk{$A_{TRRx}$, \hhs{$B_{TRRx}$, and $C_{TRRx}$}
implementations}. The other modules from vendors A and B have a smaller
\hhf{\jk{yet still a} very significant (i.e., \hhs{$>$23\%} in all cases)} fraction of
vulnerable rows. 
\hhf{We believe that modules $A0$, $A8$-$12$ are slightly more resistant to our
access pattern than the other modules of vendor A due to having more \hhs{banks}
(i.e., $16$ vs. $8$) and smaller \hhs{banks} (i.e., $32K$ vs. $64K$ rows per
bank). $B1$-$4$ have stronger rows that can endure more hammers than the other
modules of vendor B (as shown in $HC_{first}$ column of
Table~\ref{table:trr_summary}), and therefore they have a lower fraction of
vulnerable rows. $B9$-$12$ implement a different TRR version ($B_{TRR2}$),
\hhs{for} which our custom access patterns are not as effective.} 

\hhf{Modules from vendor C that implement $C_{TRR1}$ (i.e., $C0$-$8$) are less
vulnerable to our access patterns than the other modules of the same vendor. We
believe this is due to two main reasons. First, these modules use a unique row
organization that pairs every two consecutive DRAM rows, as we explain in
\S\ref{subsec:vendorC_trr}. We only observe bit flips when hammering two
aggressor rows that have odd-numbered addresses but not when the two aggressor
have even-numbered addresses. This essentially halves the number of victim rows
\hhs{where} our access patterns can cause bit flips. Second, $C0$-$6$ have
stronger rows that are less vulnerable to RowHammer than the other vendor C
modules (as Table~\ref{table:trr_summary} \hhe{shows}), and therefore $C0$-$6$
have even lower fraction of vulnerable rows than $C7$-$8$. 
}

\hhs{Overall, even though our custom RowHammer access patterns cause bit flips
in \numTestedDIMMs{} DRAM modules, we could not explore the
entire space of both TRR implementations and custom RowHammer patterns.
Therefore, we believe future work can lead to even better RowHammer patterns via
more exhaustive analysis and testing.}

\vspace{-1mm}
\subsection{Bypassing System-Level ECC \hhf{Using \method{}}}
\label{subsec:ecc_vs_rowhammer}
\vspace{-1mm}

\hhe{Although} we \hhf{clearly} show that the \hhf{custom} access patterns
we craft induce RowHammer bit flips in \hhf{a very large fraction of DRAM rows
(\S\ref{subsec:row_vulnerability})}, a system that uses Error Correction
\hhf{Codes} (ECC)~\cite{dell1997white, meza2015revisiting, schroeder2009dram,
gong2017dram, kang2014co, micron2017whitepaper, nair2016xed,
patel2019understanding, lin2001error, hamming1950error, sridharan2012study,
sridharan2015memory} can potentially protect against RowHammer bit flips if
those bit flips are distributed such that no ECC codeword contains more \hhf{bit
flips than ECC can correct}.

\begin{figure}[!h]
    \centering
    \includegraphics[width=.9\linewidth]{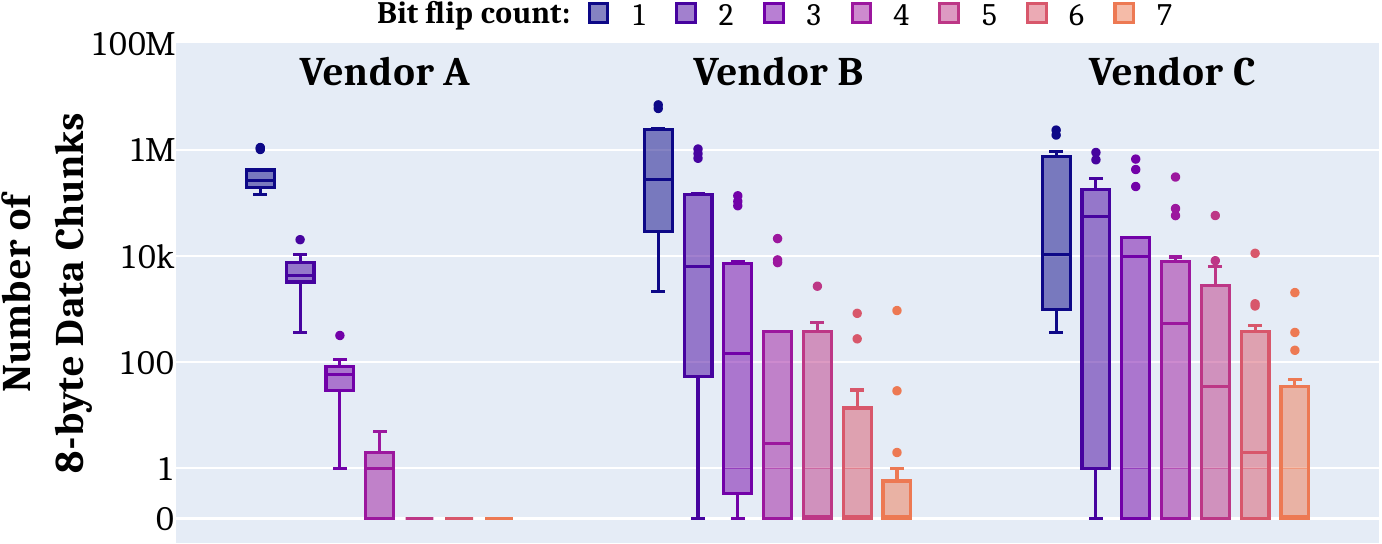}
    \vspace{-4mm}
    \caption{\hhs{Distribution} of 8-byte data chunks \hhf{(log scale) with
    different RowHammer bit \hhe{flip counts} in a single DRAM bank from \hhe{each of the}
    \numTestedDIMMs{} \hhe{tested} DDR4 modules}.}
    \vspace{-4mm}
    \label{fig:data_chunks_with_bitflips}
\end{figure}

Fig.~\ref{fig:data_chunks_with_bitflips} \hht{shows the distribution of
RowHammer bit flips \hhf{that our custom access patterns induce} \hhf{across}
8-byte data chunks as box-and-whisker plots\footnotemark[14] for \emph{all}
\hhf{\numTestedDIMMs{}} DRAM modules we test \hhf{across three vendors}.} We use
8-byte data chunks as DRAM ECC typically uses 8-byte or larger
datawords~\cite{micron2017whitepaper, oh2014a, kwak2017a, kwon2017an, im2016im,
son2015cidra, cha2017defect, jeong2020pair}.

The majority of the 8-byte data chunks are those that have only a single
\hhs{RowHammer} bit flip (i.e., up to $6.9$ million 8-byte data chunks with a
single bit flip \hhf{in one bank of} module $B13$), which can be corrected using
typical SECDED ECC~\cite{micron2017whitepaper, oh2014a, kwak2017a, kwon2017an,
im2016im, son2015cidra, cha2017defect, jeong2020pair}. However, our RowHammer
access patterns can cause \hhf{at least $3$ (up to $7$)} bit flips in \hhf{many
single datawords}, which the SECDED ECC \emph{cannot} correct \hhf{or} detect,
in all three \hhf{vendor's} modules.

\vspace{-3mm}

Chipkill~\cite{dell1997white, nair2016xed, amd2009bkdg} is a symbol-based code
conventionally designed to correct errors in one symbol (i.e., one DRAM chip
failure) and detect errors in two symbols \hhf{(i.e., two DRAM chip failures)}.
Because our access patterns cause more than two bit flips in \emph{arbitrary}
locations \hhf{(i.e., different DRAM chips)}, and thus in arbitrary symbols
within an 8-byte data chunk, Chipkill does \emph{not} provide guaranteed
protection. Reed-Solomon codes~\cite{reed1960polynomial} can be designed to
provide stronger correction/detection capability at the cost of additional
\hhf{parity-}check symbols~\cite{lin2001error, huffman2003fundamentals}. To
detect (and correct half of) the maximum number of bit flips (i.e., $7$) that
our access patterns can cause in an 8-byte data chunk, a Reed-Solomon code
\hhf{would incur} a large overhead by requiring at least $7$ parity-check
symbols~\cite{reed1960polynomial}.

We conclude that \hhf{1)} conventional DRAM ECC \emph{cannot} protect against
our \hhf{new custom} RowHammer patterns and \hhf{2)} an ECC scheme that \hhe{can
protect} \hhf{against our custom patterns requires a large number of
parity-check symbols}\hhs{, i.e., large overheads}.

%% file: sections/8_related_work.tex
\vspace{-1.5mm}
\section{Related Work}
\label{sec:related_work}
\vspace{-1.5mm}

Kim et al.~\cite{kim2014flipping} are the first to introduce and analyze the
RowHammer phenomenon. \hhe{Numerous later} works \hhe{develop} RowHammer attacks
to compromise \hhe{various} systems \hhe{in various
ways}~\cite{cojocar2019exploiting, gruss2018another, gruss2016rowhammer,
lipp2018nethammer, qiao2016new, razavi2016flip, seaborn2015exploiting,
tatar2018defeating, van2016drammer, xiao2016one, frigo2020trrespass,
ji2019pinpoint, kwong2020rambleed, mutlu2017rowhammer, van2018guardion,
kim2020revisitingrh, deridder2021smash, mutlu2019rowhammer, pessl2016drama,
bosman2016dedup, bhattacharya2016curious, jang2017sgxbomb, aga2017good,
frigo2018grand, hong2019terminal, cojocar2020we, weissman2020jackhammer,
zhang2020pthammer, rowhammergithub, yao2020deephammer} and \hhe{analyze}
RowHammer further~\cite{cojocar2020we, cojocar2019exploiting, gruss2018another,
kim2020revisitingrh, qiao2016new, tatar2018defeating, yang2019trap,
walker2021dram, orosa2021deeper, google2021halfdouble}. \hhf{To our knowledge,
this is the first work to \hhs{1)} propose an experimental methodology to
understand the inner workings of \hhs{commonly-implemented} in-DRAM
\hhs{RowHammer protection (i.e., TRR) mechanisms and 2)} use this understanding
to \hhs{create custom access patterns that} circumvent the TRR mechanisms of
modern DDR4 DRAM chips.}

\noindent \textbf{In-DRAM TRR.} \jk{We already provided extensive descriptions
of TRR and TRRespass in \S\ref{sec:intro}, \S\ref{subsec:rh_mitigation}, and
\S\ref{sec:insights}. TRRespass~\cite{frigo2020trrespass} is the most relevant
prior work to ours in understanding and circumventing TRR mechanisms, yet it is
not effective enough. While TRRespass can incur RowHammer bit flips in 13 of 42
DDR4 modules (and 5 of 13 LPDDR4 modules), TRRespass does not uncover \hhs{many}
implementation details of the TRR mechanisms, which \hhs{are} important to
circumvent TRR mechanisms. For example, in 29 out of 42 DDR4 modules (and 8 out
of 13 LPDDR4 modules), TRRespass fails to find an access pattern that can
circumvent TRR. In contrast, our new \method{} methodology can be used to
understand different aspects of a TRR mechanism \hhs{in great detail} and use
this understanding to generate specific RowHammer access patterns that
effectively incur a large number of bit flips \hhe{(as we show on
\numTestedDIMMs{} real DRAM modules)}.}

\jk{\noindent \textbf{System-level RowHammer Mitigation Techniques.} A number of
studies propose system-level RowHammer mitigation
techniques~\cite{yaglikci2021blockhammer, kim2014flipping, park2020graphene,
rh-apple, brasser2016can, konoth2018zebram, van2018guardion, aweke2016anvil,
lee2019twice, seyedzadeh2017counter, son2017making, you2019mrloc,
greenfield2016throttling, yauglikcci2021security, kim2021mithril,
taouil2021lightroad, devaux2021method}. \hhs{Recent} works~\cite{frigo2018grand,
gruss2018another, yaglikci2021blockhammer, kim2020revisitingrh} \hhs{show that
some} of these \hhs{mechanisms} are insecure, inefficient, or do not scale well
in chips \hhs{with higher vulnerability to RowHammer}. \hhs{We believe the
fundamental principles of \method{} can be useful for improving
the security of these works as well as potentially combining them with in-DRAM
TRR. We leave examining such directions to future work.}
} 

%% file: sections/9_conclusion.tex
\vspace{-1mm}
\section{Conclusion}
\label{sec:conclusion}
\vspace{-2mm}

\jk{We propose \method{}, a novel experimental methodology for
reverse-engineering the \hhs{main} RowHammer mitigation mechanism, Target Row
Refresh (TRR), implemented in modern DRAM chips. 
Using \method{}, we \hhs{1)} provide insights into \hhs{the inner workings of
existing proprietary and undocumented} TRR mechanisms and \hhs{2)} develop
\hhf{custom} DRAM access patterns to efficiently circumvent TRR in
\numTestedDIMMs{} DDR4 \hhs{DRAM} modules \hhs{from} three major vendors. We
conclude that TRR does \emph{not} provide \hhs{security} against \hhs{RowHammer}
and can be easily circumvented using the \hhs{new understanding} provided by
\method{}. We believe and hope that \method{} will facilitate future research by
enabling \hhs{rigorous and open} analysis of \hhf{RowHammer mitigation
mechanisms}\hhs{, leading to the} development of \hhs{both} new \hhs{RowHammer}
attacks and more secure RowHammer \hhf{protection} mechanisms.}